\documentclass[aps,prd,amssymb,eqsecnum,nofootinbib,floatfix, a4paper,twocolumn]{revtex4}

\usepackage{mathrsfs}
\usepackage{latexsym,amsmath,amsfonts,amssymb}
\usepackage{bm}

\usepackage{graphicx}

\allowdisplaybreaks

\newcommand{\be}{\begin{equation}}
\newcommand{\ee}{\end{equation}}
\newcommand{\bea}{\begin{eqnarray}}
\newcommand{\eea}{\end{eqnarray}}

\newcommand{\D}{\partial}
\newcommand{\E}{E_{\rm real}}
\newcommand{\Ef}{{\mathcal E}_{\rm eff}}
\newcommand{\e}{\widehat{\mathcal E}_{\rm eff}}
\newcommand{\hHf}{{{\widehat H}_{\rm eff}}}
\newcommand{\hHfS}{{{\widehat H}_{\rm Schw}}}
\newcommand{\p}{{\mathbf p}}
\newcommand{\bP}{{\mathbf P}}
\newcommand{\hQ}{\widehat Q}
\newcommand{\bu}{\bar u}
\newcommand{\br}{\bar r}
\newcommand{\hh}{ \widehat \hbar}
\newcommand{\x}{{\mathbf r}}
\newcommand{\vk}{{\mathbf k}}

\newcommand{\cM}{{\cal M}}
\newcommand{\s}{{\sigma}}
\newcommand{\zp}{{}_0p}

\begin{document}

\title{High-energy gravitational scattering and the general relativistic two-body problem}

\author{Thibault Damour}
\email{damour@ihes.fr}
\affiliation{Institut des Hautes Etudes Scientifiques, 35 route de Chartres, 91440 Bures-sur-Yvette, France}

\date{\today}

\begin{abstract}
A technique for translating the classical scattering function of two gravitationally interacting 
bodies into a corresponding (effective one-body) Hamiltonian description has been recently introduced 
[Phys.\ Rev.\ D {\bf 94}, 104015 (2016)]. Using this technique, we derive, for the first time, to
second-order in Newton's constant (i.e. one classical loop) the Hamiltonian of two point masses
having an arbitrary (possibly relativistic) relative velocity. The resulting (second post-Minkowskian) Hamiltonian
is found to have a tame high-energy structure which we relate both to gravitational self-force studies of 
large mass-ratio binary systems, and to the ultra high-energy quantum scattering results of 
Amati, Ciafaloni and Veneziano. We derive several consequences of our second post-Minkowskian Hamiltonian: (i) 
the need to use special phase-space gauges to get a tame high-energy limit; and (ii) predictions about a 
(rest-mass independent) linear Regge trajectory behavior of high-angular-momenta, high-energy circular orbits.
Ways of testing these predictions by dedicated numerical simulations are indicated.
We finally indicate a way  to connect our classical results to the quantum gravitational scattering amplitude of two
particles, and we urge amplitude experts to use their novel techniques to compute the 2-loop scattering
amplitude of scalar masses, from which one could deduce the third post-Minkowskian effective one-body Hamiltonian
\end{abstract}

\maketitle

\section{Introduction}

The recent observation 
\cite{Abbott:2016blz,Abbott:2016nmj,Abbott:2017vtc,Abbott:2017oio} of gravitational wave signals from inspiralling 
and coalescing binary black holes has been significantly helped, from the theoretical side, by the availability
of a large bank of  waveform templates, defined \cite{Taracchini:2013rva,Bohe:2016gbl}
 within the analytical effective one-body (EOB) formalism 
\cite{Buonanno:1998gg,Buonanno:2000ef,Damour:2000we,Damour:2001tu,Damour:2008gu}. 
The EOB formalism combines, in a suitably resummed format, perturbative, analytical results on the motion and
radiation of compact binaries, with some non-perturbative information extracted from numerical
simulations of coalescing black-hole binaries (see \cite{Blanchet:2013haa} for a review of perturbative results on binary
 systems, and \cite{Sperhake:2014wpa} for a review of the numerical relativity of binary black holes).
Until recently, the perturbative results used to define the EOB conservative dynamics
were mostly based on the post-Newtonian (PN) approach to the general relativistic two-body interaction. 
The conservative two-body dynamics
was derived, successively, at the second post-Newtonian (2PN) \cite{Damour1982,Damour1983}, third post-Newtonian (3PN) \cite{Damour:2001bu},
and fourth post-Newtonian (4PN) \cite{Damour:2014jta} levels (with a crucial 4PN contribution having been derived by black-hole
perturbation theory \cite{Bini:2013zaa}).  For more references on the derivation (and rederivations) of the PN-expanded dynamics, and for recent progress, 
see, \cite{Blanchet:2013haa,Jaranowski:2015lha,Marchand:2017pir}.

Anticipating on the needs of the upcoming era of high signal-to-noise-ratio gravitational-wave
observations, it is important to construct theoretically improved versions of the two-body conservative dynamics.
[Here, we consider non-spinning two-body systems of masses $m_1$, $m_2$.] With this aim in mind, a novel theoretical approach to the derivation of the general relativistic two-body interaction
(and of its EOB formulation) was recently introduced \cite{Damour:2016gwp}. The basic idea of Ref. \cite{Damour:2016gwp}
was to derive improved versions of the two-body dynamics from the (gauge-invariant) {\it scattering function} $\Phi$ linking
(half) the center of mass (c.m.) classical gravitational scattering angle $\chi$ to the total energy, ${ E}_{\rm real} \equiv \sqrt{s}$, and the total angular momentum, $J$, of the 
system\footnote{We add a subscript ``real" to the total energy to avoid confusion with our later use of a corresponding ``effective energy".
We generally use units such that $c=1$,  keeping, however, track of the factors $G \equiv G_{\rm Newton}$ and $\hbar$.}
\be
\frac12 \chi^{\rm }= \Phi(E_{\rm real}, J ; m_1, m_2, G) \,.
\ee
The (dimensionless) scattering function can be expressed as a function of  dimensionless ratios, say
\be
\frac12 \chi^{\rm }= \Phi(h, j;  \nu) \,,
\ee
where we denoted
\be \label{hEj}
h \equiv \frac{\E}{M} \, ; \  j \equiv \frac{J}{G m_1 m_2} = \frac{ J}{G \mu M}\,,
\ee
with
\be
 M \equiv m_1 +m_2;\:
 \mu \equiv \frac{m_1 m_2}{m_1+m_2};\:
 \nu  \equiv \frac{\mu}{M} = \frac{m_1 m_2}{(m_1+m_2)^2}.
\ee
As $1/j= Gm_1m_2/J$, the perturbative expansion of the (classical) scattering function in powers of the gravitational constant $G$ [{\it post-Minkowskian} (PM) expansion, 
which, contrary to the PN one does not assume slow velocities] is  seen to be equivalent to an expansion in inverse
powers of the angular momentum:
\be \label{PMexp}
\frac12 \chi^{\rm }_{\rm class}(\E, J) = \frac1j \chi_{1}(h, \nu) + \frac1{j^2} \chi_{2}(h, \nu) + \frac1{j^3} \chi_{3}(h, \nu) + \cdots
\ee
Here, $\chi_{1}(h, \nu)/j$  is the first post-Minkowskian (1PM) approximation of (half) the scattering function, 
$\chi_{2}(h, \nu)/j^2$   the second post-Minkowskian (2PM) one, $\chi_{3}(h, \nu)/j^3$   the third post-Minkowskian (3PM) one, etc.

Ref. \cite{Damour:2016gwp} (re)derived the leading-order (LO), 1PM approximation $\chi_{1}(h, \nu)/j$  to the scattering function
(first derived in \cite{Portilla:1980uz}), emphasized its link to the corresponding LO quantum scattering amplitude, and showed how to transcribe it
within EOB theory. [The latter transformation is crucial for translating an information valid for hyperboliclike motions (scattering states)
into an information concerning ellipticlike motions (bound states), as most relevant for gravitational-wave physics.] 
The generalization of 1PM scattering to spinning bodies has been recently worked out 
\cite{Bini:2017xzy,Vines:2017hyw}.

The first aim of
the present work will be to extend the results of Ref. \cite{Damour:2016gwp} to the next-to-leading order (NLO) in the expansion
in powers of $G$, i.e. to the 2PM level ($O(G^2)$). This will be done by using the 2PM-level results derived more than thirty years ago in 
Refs.~\cite{Westpfahl:1979gu,Bel:1981be,Westpfahl:1985,Westpfahl:1987}.  As we shall discuss below, the EOB transcription of
the 2PM-level scattering $\chi_{1}(h, \nu)/j + \chi_{2}(h, \nu)/j^2$ yields interesting insights on the {\it high-energy behavior} of
the gravitational interaction, and of its EOB formulation.

The second aim of the present work will be to usher in techniques for translating (via an EOB formulation)
{\it quantum gravitational scattering results} into quantities of direct use for improving the description of the
classical dynamics of  compact binaries (such as inspiralling and coalescing binary black holes).
There has been many advances in perturbative quantum gravity (and notably high-energy scattering), 
coming from various avenues, notably: string theory \cite{Amati:1987wq,Amati:1990xe,Giddings:2007bw},
eikonal quantum field theory \cite{tHooft:1987vrq,Kabat:1992tb,Akhoury:2013yua}, gauge-gravity duality
\cite{Kawai:1985xq,Bern:2008qj,Bern:2010ue,Cachazo:2013iea,Bern:2017yxu}, and on-shell techniques \cite{Bjerrum-Bohr:2013bxa,Bjerrum-Bohr:2014zsa,Bjerrum-Bohr:2016hpa}.
We shall make contact with some of these results (notably the high-energy scattering results of Amati, Ciafaloni and Veneziano
\cite{Amati:1987wq,Amati:1990xe}), and indicate what would be the most interesting quantum scattering
amplitudes to compute to significantly improve our knowledge of the general relativistic dynamics of two-body systems.

\section{Classical two-body scattering function at next to leading order (second post-Minkowskian approximation).} \label{sec2}

The (classical) relativistic gravitational two-body scattering function $\frac12 \chi_{\rm classical}(  {E}_{\rm real}, J; m_1, m_2;G)$ 
can be obtained as a power series in $G$ by iteratively solving the  equations of motion of the two worldlines, together
with Einstein's gravitational field equations. Let us sketch here how the computation (in PM perturbation theory) of
the scattering function can be naturally represented as a sum of Feynman-like diagrams. The main purpose of the present
Section is to exhibit the similarity of the the latter classical scattering diagrams to the usual,
quantum (Feynman) diagrams. It would be interesting to study whether this similarity would allow one to import, or
translate, the improved, modern computational quantum amplitude techniques mentioned above into 
corresponding, improved classical scattering computations.

The equation of motion of each worldline $x_a^{\mu}=x_a^{\mu}(\s_a)$ (with $a=1,2$)
can be written (in first-order form) as the Euler-Lagrange equations of the Hamiltonian ${\mathcal H}= \frac12 g^{\alpha \beta}(x) p_{\alpha} p_{\beta}$, namely
\begin{align} \label{eoma}
&\frac{d x_a^{ \mu}}{d \sigma_a}=  g^{\mu \nu}(x_a) p_{a \nu}\,, \nonumber \\
&\frac{d p_{a \mu}}{d \sigma_a} = -\frac12 \,\D_\mu g^{\alpha \beta}(x_a) \, p_{a \alpha} p_{a \beta}\,.
\end{align}
We use a mostly positive signature with, e.g., $ g^{\mu \nu} p_{a \mu} p_{a \nu} = - m_a^2$. Each worldline parameter
$\s_a$ is linked to the corresponding proper time $s_a$ via $\sigma_a= s_a/m_a$. 

The equations of motion of the (contravariant)
metric $ g^{\mu \nu}(x)$ are obtained by gauge-fixing the Einstein equations $R^{\mu \nu}-\frac12 R g^{\mu \nu}= 8\pi G T^{\mu \nu}$.
Using, say, harmonic coordinates, one gets equations for $g^{\mu \nu}(x) \equiv \eta^{\mu \nu} - h^{\mu \nu}(x)$ of the form (in four spacetime dimensions)
\be \label{eomh}
\Box h^{\mu \nu} = - 16 \pi G S^{\mu \nu} + O(\D \D h h + hS)\,,
\ee
with $\Box= \eta^{\mu \nu} \D_{\mu} \D_{ \nu}$,
\be
S^{\mu \nu}= T^{\mu \nu} - \frac1{2} T^{\alpha \beta} g_{\alpha \beta} g^{\mu \nu} \,,
\ee
 and 
\be
T^{\mu \nu}(x)= \sum_{a=1,2} \int d\s_a   p_a^{\mu} p_a^{\nu} \, \frac{\delta^4(x-x_a(\sigma_a))}{\sqrt{g}}\,,
\ee
where $ p_a^{\mu} \equiv g^{\mu \nu} p_{a \nu}$ and $g=- \det g_{\mu \nu}$.

The scattering function is obtained from the total change of the 4-momenta between the infinite past
and the infinite future:
\begin{align} \label{deltapmu}
\Delta p_{a \mu} &= \int_{- \infty}^{+\infty} d \sigma_a \frac{d p_{a \mu}}{d \sigma_a} \nonumber \\
&= -\frac12 \int_{- \infty}^{+\infty} d \sigma_a \,\D_\mu g^{\alpha \beta}(x_a) \, p_{a \alpha} p_{a \beta}\,.
\end{align}
More precisely, the (absolute value of the)  scattering angle $\chi$ in the center of mass (c.m.) frame
 is related to the magnitude of the spatial projections (in the c.m. frame)
$\Delta{\bf p}= \Delta {\bf p}_1 = - \Delta {\bf p}_2$ of $\Delta p_{a \mu}$\footnote{We assume that 
we are solving the two-body problem with the time-symmetric
Green's function ${\mathcal G}_{\rm sym} = \frac12 ({\mathcal G}_{\rm ret}+ {\mathcal G}_{\rm ad})$, so that the
dynamics is conservative.} via 
\be \label{sinchi/2}
\sin \frac{\chi}{2}=\frac{ |\Delta {\bf p}_a|}{2 \,  |{\bf p}_a|} = \frac{ |\Delta {\bf p}_a|}{2 \, {p}_{\rm c.m.}}\,.
\ee

The integral expression \eqref{deltapmu} can  be used as the basis of a perturbative computation of $\chi$.
If we start by considering the two worldlines as given, we can insert in  \eqref{deltapmu} the iterative solution of the field
equations \eqref{eomh}, say [with a (time-symmetric) Green's function ${\mathcal G}(x-y)$ satisfying $\Box {\mathcal G}(x-y)= - 4 \pi \delta^4(x-y)$]
\be \label{solh}
h^{\mu \nu}(x) = 4 G \int d^4y \, {\mathcal G}(x-y)   S^{\mu \nu}(y) + O(G^2)\,.
\ee

\begin{figure}
\includegraphics[scale=1.1]{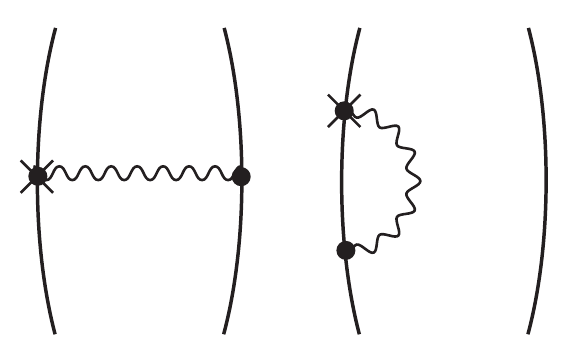}
\caption{\label{fig1}
Feynman-like diagrams for the {\it classical} gravitational scattering at first order in $G$.}
\end{figure}

At LO in $G$ this yields the following integral expression for $\Delta p_{1 \mu}$ 
\begin{align} \label{deltap1G1}
&\Delta  p_{1 \mu} = 2 G \int d\s_1 d\s_2 p_{1 \alpha} p_{1 \beta} \nonumber \\
&\times \D_{\mu} {\mathcal P}^{\alpha \beta ;\alpha' \beta'} (x_1(\s_1) -x_2(\s_2))  p_{2 \alpha'} p_{2 \beta'}  \nonumber \\
 +2 G &\int d\s_1 d\s'_1 p_{1 \alpha} p_{1 \beta} \D_{\mu} {\mathcal P}^{\alpha \beta ;\alpha' \beta'} (x_1(\s_1) -x_1(\s'_1))  p_{1 \alpha'} p_{1 \beta'}  \nonumber \\
&+O(G^2)
\end{align}
where 
\be \label{gravitonpropagator}
{\mathcal P}^{\alpha \beta ;\alpha' \beta'}(x-y) = \left(\eta^{\alpha \alpha'} \eta^{\beta \beta'} -\frac12 \eta^{\alpha \beta}  \eta^{\alpha \beta'}\right) {\mathcal G}(x-y),
\ee
denotes the graviton propagator (in $x$ space). 

It is natural to associate with the two $O(G^1)$  terms in Eq. \eqref{deltap1G1}
the two Feynman-like diagrams of Fig. 1. The crosses on the left worldline [corresponding to $x_1(\s_1)$] represent 
the partial derivatives $\D_{\mu}$ acting on the graviton propagators. The diagram on the left correspond to the first
integral on the right-hand side (rhs) of Eq. \eqref{deltap1G1} (involving a propagation of the gravitational interaction between the
two worldlines), while the diagram on the right correspond to the second integral (involving a ``gravitational loop",
i.e. a propagation of the gravitational interaction between the same worldline $x_1(\s_1)$). More about this below.

At second order in the iterative solution of the field equation \eqref{eomh} (still assuming some given worldlines
$x_a^{\mu}(\sigma_a), p_{a \mu}(\sigma_a)$), there will be further contributions, of order at least
$O(G^2)$, to $\Delta  p_{1 \mu}$, some of whose diagrammatic representations are illustrated in Fig. 2. [The cubic vertices,
between the two worldlines, in Fig. 2 represent the cubically nonlinear gravitational interactions. The wiggly lines represent,
as in Fig. 1, the graviton propagator \eqref{gravitonpropagator}.]

\begin{figure}
\includegraphics[scale=1.1]{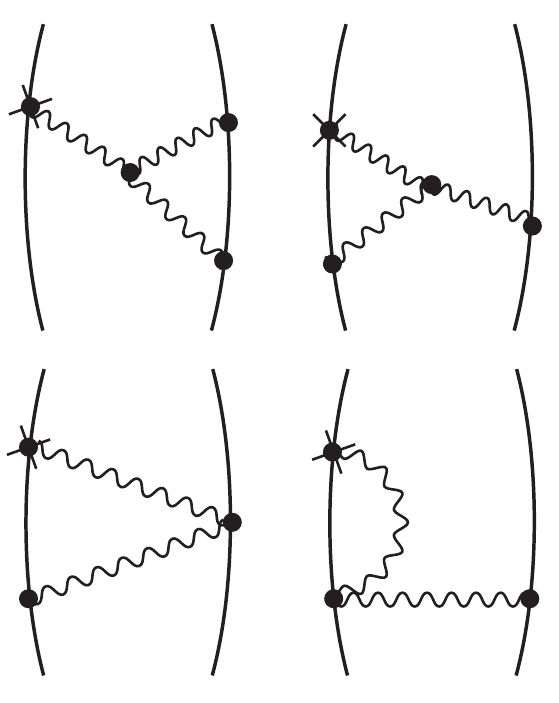}
\caption{\label{fig1}
Some of the Feynman-like diagrams for the {\it classical} gravitational scattering at second order in $G$.}
\end{figure}

However, this is not the complete story because the above integral expressions for $\Delta  p_{1 \mu}$ (graphically
represented in Fig. 1 and Fig. 2) had assumed that the worldlines $x_a^{\mu}(\sigma_a), p_{a \mu}(\sigma_a)$ were
some given solutions of the (interacting) equations of motion \eqref{eoma}. [This is why they have been drawn
as {\it curved} worldlines in Fig. 1 and Fig. 2.]
To convert the latter  formal perturbative expansion for \eqref{deltapmu} into an {\it explicit} perturbative
series for the scattering function, one needs to complement it by a perturbative expansion of the worldlines
themselves: 
\begin{align}
x_a^{\mu}(\sigma_a) &={}_0x_a^{\mu}(\sigma_a) + G\, {}_1x_a^{\mu}(\sigma_a) + G^2\, {}_2x_a^{\mu}(\sigma_a)+\cdots, \nonumber \\
 p_{a \mu}(\sigma_a) &={}_0p_{a \mu}(\sigma_a) + G \,{}_1p_{a \mu}(\sigma_a) + G^2 \,{}_2p_{a \mu}(\sigma_a)+\cdots 
\end{align}

The LO ($O(G^0)$)  worldline solution is a set of two straight worldlines, say ${}_0x_a^{\mu}(\sigma_a)= {}_0x_a^{\mu}(0) + {\zp}_a^{\mu} \, \sigma_a $, where ${\zp}_a^{\mu}$ are constant momenta (say the  incoming 4-momenta of the particles).
Inserting this LO worldline solution in the perturbative solution \eqref{solh} of the field equation, and neglecting $O(G^2)$ 
corrections, yields an explicit  1PM ($O(G^1)$) metric perturbation
\begin{align} 
\label{h1}
 G \,{}_1h^{\mu \nu}(x) &= 4 G \int d^4y \, {\mathcal G}(x-y)   
 \nonumber \\
 &\times \left({}_0T^{\mu \nu}(y) - \frac1{2} {}_0T^{\alpha \beta}(y) \eta_{\alpha \beta} \eta^{\mu \nu}  \right) \,,
\end{align}
where
\be
{}_0T^{\mu \nu}(x)= \sum_{a=1,2} \int d\s_a  \, {\zp}_a^{\mu} \, {\zp}_a^{\nu} \, \delta^4(x-{}_0x_a(\sigma_a)).
\ee
Inserting then the  $O(G^1)$ solution  \eqref{h1} in the
worldline equations of motion \eqref{eoma}, yields $O(G^1)$ worldline equations of motion for 
$x_a^{\mu}(\sigma_a)= {}_0x_a^{\mu}(\sigma_a) + G {}_1x_a^{\mu}(\sigma_a) + O(G^2)$, namely
\begin{align} \label{eomaG}
&\frac{d {}_1x_a^{ \mu}}{d \sigma_a}=  \eta^{\mu \nu} {}_1p_{a \nu} - {}_1h^{\mu \nu}(x_a) \,{}_0p_{a \nu}, \nonumber \\
&\frac{d {}_1p_{a \mu}}{d \sigma_a} = \frac12 \,\D_\mu \, {}_1h^{\alpha \beta}(x_a) \, {}_0p_{a \alpha} \, {}_0p_{a \beta}\,.
\end{align}
The $O(G^1)$ correction ${}_1p_{a \mu}(\sigma_a)$ to the momenta $p_{a \mu}(\sigma_a) = {}_0p_{a \mu}(\sigma_a)+ G \, {}_1p_{a \mu}(\sigma_a) + O(G^2)$
(with the boundary condition that $\lim_{- \infty} p_{a \mu}(\sigma_a)= {}_0p_{a \mu}$) is  obtained as an integral, namely
\be \label{p1sigma}
{}_1p_{a \mu}(\sigma_a) = \int_{- \infty}^{\sigma_a} d \sigma'_a \frac12 \,\D_\mu \, {}_1h^{\alpha \beta}(x_a(\sigma'_a)) \, {}_0p_{a \alpha} \, {}_0p_{a \beta}.
\ee
Then the $O(G^1)$ correction ${}_1x_a^{\mu}(\sigma_a)$ to the worldline $x_a^{\mu}(\sigma_a)= {}_0x_a^{\mu}(\sigma_a) + G \,{}_1x_a^{\mu}(\sigma_a) + O(G^2)$  is obtained by integrating the first equation \eqref{eomaG} with suitable boundary
conditions in the infinite past. [Because of the $\sim 1/\sigma_a$ decrease of the rhs of the first equation \eqref{eomaG} one must separate a logarithmic term before imposing a usual decaying boundary condition at $\s_a \to - \infty$.]

The explicit, first-post-Minkowskian (1PM) [$O(G^1)$] value of the  scattering angle is then obtained by computing the $\s_a \to + \infty$ limit
of ${}_1p_{a \mu}(\sigma_a)$ and inserting it in  Eq. \eqref{sinchi/2}. The explicit integral expression of  $\Delta p_{a \mu}$ defined
by the $\s_a \to + \infty$ limit of \eqref{p1sigma} is obtained from the previous result \eqref{deltap1G1} by replacing everywhere
on the rhs $x_a^{\mu}(\sigma_a)$ by ${}_0x_a^{\mu}(\sigma_a)= {}_0x_a^{\mu}(0) + {\zp}_a^{\mu} \, \sigma_a $,
and $p_{a \mu}(\sigma_a)$ by ${\zp}_a^{\mu}$, where we recall that ${\zp}_a^{\mu}$ are the constant, incoming momenta of the particles.
The latter explicit integral expression for $\Delta p_{a \mu}$, at the 1PM order, can be 
vizualized by diagrams similar to those of Fig. 1, except for the fact that the two worldlines must now be
drawn as {\it straight} worldlines $x_a^{\mu}(\sigma_a)= {}_0x_a^{\mu}(\sigma_a)= {}_0x_a^{\mu}(0) + {\zp}_a^{\mu} \, \sigma_a $.
It is then shown that (after regularization) the
second (one-loop) diagram in Fig. 1 gives a vanishing contribution, so that the 1PM scattering angle is proportional to $G m_1 m_2$,
and obtainable from the single explicit integral (which no longer assumes that the worldlines are known beforehand)
\begin{align} \label{deltap1PM}
&\Delta  p_{1 \mu} = 2 G \int d\s_1 d\s_2 \,  {\zp}_{1 \alpha} \, {\zp}_{1 \beta} \nonumber \\
&\times \D_{\mu} {\mathcal P}^{\alpha \beta ;\alpha' \beta'} ({}_0x_1(\s_1) -{}_0x_2(\s_2)) \, {\zp}_{2 \alpha'}\, {\zp}_{2 \beta'}  \nonumber \\
&+O(G^2) \,.
\end{align}

The 1PM integral \eqref{deltap1PM} can either be computed in ${\bm x}$-space (using the simple, ${\bm x}$-space value
of the scalar Green's function ${\mathcal G}(x-y)= \delta[(x-y)^2]$), as was done long ago in Ref.\cite{Portilla:1980uz},
or in the Fourier domain (using ${\mathcal G}(k)= 4\pi/k^2$), as was recently done in Ref. \cite{Damour:2016gwp}. The
explicit, final result for the $O(G^1)$classical  scattering angle is
\be \label{chiG}
\sin \frac{\chi_{\rm class}^{O(G)}}{2}= \frac{ G }{ J }  \frac{ 2 (p_1.p_2)^2 - p_1^2 p_2^2 }{ \sqrt{(p_1.p_2)^2 - p_1^2 p_2^2} }\,.
\ee
It was emphasized in \cite{Damour:2016gwp} that the Fourier-domain computation of $\Delta  p_{1 \mu}$ closely parallels
the computation of the corresponding LO, one-graviton-exchange,  Feynman gravitational scattering amplitude (described
by a Feynman diagram similar to the left diagram in Fig. 1). The quantum scattering amplitude
 ${\mathcal M}$ for the scattering of massive scalar particles, reads (see, e.g. \cite{Kabat:1992tb})
 \be
 {\mathcal M}^{O(G)}(s,t)= 16 \pi \frac{G}{\hbar} \frac{2 (p_1.p_2)^2 - p_1^2 p_2^2}{-t}\,,
 \ee
 where $s \equiv - (p_1+p_2)^2= E_{\rm real \, c.m.}^2$, and 
 $-t \equiv (p'_1-p_1)^2= {\bf q}^2_{\rm c.m.}= 4 {p}_{\rm c.m.}^2 \sin^2 \frac{\chi}{2}$,
 are the usual Mandelstam quantities. [See below for more explanations about the definition, and sign convention,
 for $\cM$.]
We recall that ${E}_{\rm real}$, and $J$, are both evaluated in the {\it c.m. frame} of the two-body system.

At the order $G^2$, i.e at the second post-Minkowskian (2PM) order, one can write down an explicit integral
expression for $\Delta  p_{1 \mu}$ by inserting the next-to-leading-order (NLO), $O(G^1)$, solutions for the
worldlines (obtained, as explained above, by inserting Eq. \eqref{p1sigma} in the first equation \eqref{eomaG})
in the general iterative expression \eqref{deltap1G1}, and its $O(G^2)$ analog (partly graphically represented in Fig. 2).
When aiming at the 2PM accuracy, it is enough to replace in the $O(G^2)$ diagrams of Fig. 2 the curved worldlines by
the LO straight worldlines ${}_0x_a^{\mu}(\sigma_a)= {}_0x_a^{\mu}(0) + {\zp}_a^{\mu} \, \sigma_a $, where ${\zp}_a^{\mu}$.
However, one must insert in the formally $O(G^1)$ diagrams of Fig. 1 the NLO, $O(G^1)$, solutions for the worldlines,
i.e. Eq. \eqref{p1sigma} for $p_{a \mu}(\sigma_a) $, and the corresponding, explicit  $O(G^1)$ solution for
$x_a^{\mu}(\sigma_a)= {}_0x_a^{\mu}(\sigma_a) + G {}_1x_a^{\mu}(\sigma_a) + O(G^2)$, involving a {\it double} integral expression for 
${}_1x_a^{\mu}(\sigma_a)$.  The corresponding extra $O(G^2)$ contributions to the classical scattering $\Delta  p_{1 \mu}$
can be vizualized as additional  $O(G^2)$ diagrams of the ladder (and crossed-ladder) type, which are the
classical analogs of  the usual quantum ladder diagrams. [One can check that the classical ladder diagrams contain (in Fourier-space) the 
denominators $\sim 1/(k.p_a)$ that are resummed in the eikonal-approximation to the quantum ladder diagrams.]

The so-obtained explicit, 2PM [$O(G^2)$] value of the  scattering angle has been computed in Refs. \cite{Westpfahl:1979gu,Westpfahl:1985,Westpfahl:1987}, using the explicit ${\bm x}$-space 2PM ($O(G^2)$) equations of motion \cite{Westpfahl:1979gu,Bel:1981be}.
It can be written as\footnote{To the 2PM accuracy considered here, one could equivalently write the lhs of \eqref{chiPM1+2} as $\sin \frac12 \chi$.}
\be \label{chiPM1+2}
\frac12 \chi^{\rm }_{\rm class}(\E, J) = \frac1j \chi_{1}(\e, \nu) + \frac1{j^2} \chi_{2}(\e, \nu) + O(G^3),
\ee
 where
\be \label{chi1PM}
\chi_{1}(\e, \nu) = \frac{2 \, \e^2 -1}{\sqrt{\e^2-1}} \, ,
\ee
and
\be \label{chi2PM}
\chi_{2}(\e, \nu) = \frac{3 \pi}{8}\, \frac{5 \, \e^2 -1}{\sqrt{1+ 2\nu (\e-1)}} \, .
\ee
Here, we have replaced the total (c.m.) angular momentum $J$ by its dimensionless counterpart  $ j \equiv \frac{J}{G m_1 m_2}$,
and the total c.m. energy  $E_{\rm real} = \sqrt{s}$ by the dimensionless energy variable
\be \label{f}
\e \equiv \frac{{\mathcal E}_{\rm eff}}{\mu} \equiv  \frac{({E}_{\rm real})^2 - m_1^2  -m_2^2 }{2 \, m_1 m_2} =  \frac{s - m_1^2  -m_2^2 }{2 \, m_1 m_2}.
\ee
The ``effective energy" ${\mathcal E}_{\rm eff} = \mu \e $ plays a central role in EOB theory \cite{Buonanno:2000ef,Damour:2000we},
and the map $f$ between ${\mathcal E}_{\rm real}$ and ${\mathcal E}_{\rm eff}$ defined by Eq.  \eqref{f} was recently shown \cite{Damour:2016gwp} to be exact to all orders in the PN expansion.  
Note again that the expansion in powers of $1/j=Gm_1m_2/J$ in \eqref{chiPM1+2} is equivalent to the PM expansion
in powers of $G$.

The extreme mass-ratio limit $m_1 \ll m_2$, corresponds to  $\nu \ll1$. In this limit, the scattering angle $\chi$ should reduce
to the scattering angle of a test-particle moving around a Schwarzschild black hole of mass $M=m_1+m_2 \approx m_2$.
We shall check in the next section that this is indeed the case. In the mean time, note that the $O(G)$  energy-dependent coefficient 
$\chi_{1}(\e, \nu)$ is actually independent of the symmetric mass-ratio $\nu$, while the $O(G^2)$  coefficient 
$\chi_{2}(\e, \nu)$ depends on $\nu$ only through the factor (remembering the definition \eqref{hEj})
\be \label{1byh}
\frac{1}{\sqrt{1+ 2\nu (\e-1)}}= \frac1h = \frac{M}{\E} \,,
\ee
multiplying the $O(G^2)$ test-particle ($\nu \to 0$) result
\be
\chi_{2}^{\rm Schwarz}(\e) = \frac{3 \pi}{8}\, (5 \, \e^2 -1)\,.
\ee
In Eq. \eqref{1byh}, we used the well-known EOB inverse energy map, namely 
\be \label{finv}
\E= M \sqrt{1+ 2\nu (\e-1)} , 
\ee
which inverts the quadratic
relation ${\mathcal E}_{\rm eff}= f({\mathcal E}_{\rm real})$, Eq. \eqref{f}. 

Another check of the 2PM scattering angle can be obtained by reexpanding the $G$ expansion  \eqref{chiPM1+2}
(each term of which is an exact function of the energy) in a PN way, i.e. in powers of $ 1/c^2$. This can be done,
for instance, by decomposing $\E$ in rest-mass plus non-relativistic energy, say
\be
\E = M c^2 + \frac12 \mu v_E^2.
\ee
[Here, we are not making an approximation but simply introducing the notation $v_E$  for $\sqrt{ 2(\E-Mc^2)/\mu}$.]
The expansion in powers of  $ v_E/c$ then corresponds to a PN expansion. When doing
so, one must also remember that the exact definition of the dimensionless
angular momentum $j$ involves one power of $c$: $j \equiv c J/(Gm_1 m_2)$.
One can then check the values
of the coefficients $\chi_{m n}(\nu) $ (with $m=1,2$ and $2(m-1)+n \leq 8$)
in the 4PN-level ($O(1/c^8)$) expansion of \eqref{chiPM1+2}, say
\bea \label{chiPN0}
 && \frac{1}{2} \chi^{}(E, J)   = \frac{Gm_1 m_2}{v_E J} \Bigl(1+ \chi_{12}(\nu) \left( \frac{v_E}{c}\right)^2 \nonumber\\
&& \qquad + \chi_{14}(\nu) \left(\frac{v_E}{c}\right)^4 + \cdots +\chi_{18}(\nu) \left(\frac{v_E}{c}\right)^8 \Bigl) \nonumber\\
& & \qquad + \left( \frac{G m_1 m_2}{c J} \right)^2  \Bigl(1 + \chi_{22}(\nu) \left(\frac{v_E}{c}\right)^2 \nonumber \\
&& \qquad + \chi_{24}(\nu) \left(\frac{v_E}{c}\right)^4 + \chi_{26}(\nu) \left(\frac{v_E}{c}\right)^6\Bigl) 
\nonumber \\
&& \qquad + O\left(\frac{1}{c^{10}}\right)+ O\left( \left(\frac{G m_1 m_2}{ J} \right)^3 \right) \, ,
\eea
against the explicit PN-expanded scattering results of Refs. \cite{Bini:2012ji,Bini:2017wfr}.
[Let us note again, in this respect, a typo in the first Eq. (5.51) (defining $A_{2a}$) in \cite{Bini:2012ji}:
the sign of the coefficient of $\tilde E$ on the rhs should be reversed, namely it should read $+ (5-2\nu)/2$.]
Note also that Ref. \cite{Bini:2017wfr} use PN expansions based on the different 
decomposition $\e^2=1+ v_{\infty}^2/c^2$.

\section{Classical  scattering angle of a test-particle around a Schwarzschild black hole.} 
\label{secSchw}

As a check on the 2PM result above (obtained in Refs. \cite{Westpfahl:1979gu,Westpfahl:1985,Westpfahl:1987}), and as
a warm up towards its EOB transcription, let us consider the scattering of a test-particle of mass $m_0$ around a
Schwarzschild black hole of mass $M_0$. [Below we shall identify $m_0$ to $\mu$, and $M_0$ to $M$.]
As usual, we start from the mass-shell condition
\be \label{Schwamassshell}
0= g_0^{\mu \nu} P_{\mu} P_{\nu} + m_0^2\,,
\ee
in the Schwarzschild metric 
\be \label{g0}
 g^0_{\,\mu \nu} dx^{\mu} dx^{\nu} = - A_0 dT^2 + B_0 dR^2 + C_0 (d \theta^2 + \sin^2 \theta d \varphi^2),
\ee
where
\be \label{A0B0C0}
A_0(R)= 1- \frac{2 G M_0}{R} \, ; \, B_0(R)= \frac{1}{1- \frac{2 G M_0}{R}} \, ; \,  C_0(R)= R^2\,.
\ee
The simplest way to compute the scattering angle is to use Hamilton-Jacobi theory, i.e. to replace $P_{\mu}=\D_{\mu} S_{0}$
in the mass-shell condition \eqref{Schwamassshell}
and to solve for the radial action $S^{0}_ R$ (considered for equatorial motions) in
\be
S_{0}= - {\mathcal E}_0 T +  P_{\varphi} \varphi + S^{0}_R(R ; {\mathcal E}_0, P_{\varphi})\,.
\ee
This yields
\be
S^{0}_R(R  ; {\mathcal E}_0, P_{\varphi})) = \int dR P_R(R ; {\mathcal E}_0, P_{\varphi})
\ee
where
\be
P_R(R ; {\mathcal E}_0, P_{\varphi}) = \pm  \sqrt{\frac{B_0}{A_0}  } \sqrt{ {\mathcal E}_0^2 - A_0 \left(m_0^2 + \frac{P_{\varphi}^2}{C_0}\right) }.
\ee
The relation between the angle $\varphi$ and the radial coordinate $R$ is then obtained from  (with $J_0 \equiv P_{\varphi}$)
\be
\varphi(R) = - \int dR \frac{\partial P_R(R ; {\mathcal E}_0, J_0)}{\partial J_0}  \,,
\ee
which yields
\be
\varphi(R) = J_0 \int \frac{dR}{C_0} \frac{\sqrt{A_0 B_0} }{\pm \sqrt{{\mathcal E}_0^2 -A_0 \left(m_0^2 + \frac{J_0^2}{C_0} \right) } }.
\ee
The scattering angle $\chi$ is then obtained by subtracting $\pi$ from the full, two-sided radial integral, taken from the
incoming state (at time $ -\infty$, i.e. $R= + \infty$ and a negative sign for the squareroot) to the final state (at time
$ +\infty$, i.e. $R= + \infty$ and a positive sign for the squareroot):
\be \label{chigen1}
\pi + \chi=  \int_{-\infty}^{+\infty} J_0\frac{dR}{C_0} \frac{\sqrt{A_0 B_0} }{\pm \sqrt{{\mathcal E}_0^2 -A_0 \left(m_0^2 + \frac{J_0^2}{C_0} \right) } } \, .
\ee
This expression (which is valid for any metric of the form \eqref{g0}) simplifies in the case of the Schwarzschild metric \eqref{A0B0C0}
(for which, in particular $A_0B_0=1$ and $C_0=R^2$).  It is convenient to replace the original variables $R,  {\mathcal E}_0, J_0$
by the corresponding rescaled, dimensionless variables  $r,  \widehat{\mathcal E}_0, j_0$ defined as
\be
R \equiv G M_0 \, r \, ; \, {\mathcal E}_0 \equiv m_0 \, \widehat{\mathcal E}_0 \, ; \, J_0 \equiv G M_0 m_0 \, j_0 \, .
\ee
Introducing also the dimensionless integration variable
\be
y \equiv \frac{j_0}{r} = \frac{J_0}{m_0 R}\,,
\ee
we get
\be
\frac{\pi}{2} + \frac{\chi}{2} = \int_0^{y_0^{\rm max}} \frac{dy}{\sqrt{\widehat{\mathcal E}_0^2 -\left(1- \frac{2}{j_0} y \right) (1+y^2)}}
\ee
where $y_0^{\rm max}$ (which depends on $\widehat{\mathcal E}_0$ and $j_0$) denotes the positive root of the radical that is closest to $0$.
With the further notation
\be \label{c0}
c_0 \equiv \widehat{\mathcal E}_0^2 -1\,,
\ee
we have
\be \label{schwaint}
\frac{\chi}{2}=  \int_0^{y_0^{\rm max}(c_0, j_0)} \frac{dy}{\sqrt{c_0 - y^2  + \frac{2}{j_0} y (1+y^2)}} - \frac{\pi}{2}\,.
\ee
The latter integral expression is convenient for expanding $\chi$ in powers of $1/j_0$, i.e. for computing the coefficients in
the PM expansion \eqref{PMexp} of the Schwarzschild scattering angle.  When $ j_0 \to \infty$ (so that $y_0^{\rm max} \to \sqrt{c_0}$),
the integral on the rhs becomes $\int_0^{\sqrt{c_0}} dy/\sqrt{c_0 - y^2}= \frac{\pi}{2}$, yielding, as needed, $\lim_{j_0 \to \infty} \chi=0$.
The successive terms $\chi_1/j_0 + \chi_2/j_0^2 +\cdots$ in the PM expansion \eqref{PMexp} can then be computed from 
the expansion of the integrand $1/\sqrt{c_0 - y^2  + \frac{2}{j_0} y (1+y^2)}$ in successive powers of $1/j_0$. Actually, there are two
subtleties linked to this expansion. On the one hand, the upper limit $y_0^{\rm max}$ of the integral also depends on $j_0$. On the
other hand, the formal expansion of $1/\sqrt{c_0 - y^2  + \frac{2}{j_0} y (1+y^2)}$ in powers of $1/j_0$, say [denoting $N_n \equiv {-\frac{1}{2} \choose  n }$]
\be
 \frac{1}{\sqrt{c_0 - y^2  + \frac{2}{j_0} y (1+y^2)}}= \sum_{n \geq0} \frac{N_n}{j_0^n} \frac{\left(2 \, y (1+y^2) \right)^n}{(c_0 - y^2)^{ \frac{2n+1}{2}}},
\ee
 involves denominators  $ \sim (c_0 - y^2)^{\frac{2n+1}{2}}$ that become increasingly singular near the upper limit of the integral. It was shown in 
Ref. \cite{Damour:1988mr} that the correct values of the coefficients in the $1/j_0$ expansion of integrals of the type \eqref{schwaint}
is very simply obtained by taking the Hadamard partie finie (Pf) of the singular integrals generated by the expansion above, i.e.
\be
\frac{\chi}{2}= \sum_{n \geq1} \frac{N_n}{j_0^n} {\rm Pf} \int_0^{\sqrt{c_0}} dy\, \frac{\left(2 y (1+y^2) \right)^n}{(c_0 - y^2)^{ \frac{2n+1}{2}}}\,.
\ee
Computing the latter Hadamard-regularized integrals yields, for the coefficients in the PM expansion of the Schwarzschild
scattering angle (using the notation \eqref{c0})
\be
\frac12 {\chi}^{\rm Schw}(\widehat{\mathcal E}_0, j_0) = \sum_{n\geq1} \frac{\chi_n^{\rm Schw}(c_0)}{j_0^n},
\ee
where
\begin{align} 
\label{chischw1}
& \chi_1^{\rm Schw}(c_0) = \frac{2 \, c_0+1}{\sqrt{c_0}}= \frac{2 \, \widehat{\mathcal E}_0^2-1 }{\sqrt{\widehat{\mathcal E}_0^2-1}},\\
\label{chischw2}
&\chi_2^{\rm Schw}(c_0)= \frac{3 \pi}{8} (5 \, c_0+4)= \frac{3 \pi}{8} (5 \,\widehat{\mathcal E}_0^2-1), \\
&\chi_3^{\rm Schw}(c_0)=\frac{64\, c_0^3 + 72 \,c_0^2 + 12\, c_0 -1}{3 \, c_0^{3/2}}, \\
&\chi_4^{\rm Schw}(c_0) = \frac{105 \pi}{128} (33 \, c_0^2 + 48 \, c_0+16).
\end{align}

It is then easily seen that  the test-mass limit ($\nu \to 0$) of both the 1PM-accurate, Eq. \eqref{chi1PM}, and the 2PM-accurate, 
Eq. \eqref{chi2PM}, scattering angles agree with the corresponding Schwarzschild results, Eqs. \eqref{chischw1}, \eqref{chischw2},
under the identifications $m_0 = \mu$, $M_0=M$, $j_0=j$, and $\widehat{\mathcal E}_0 = \e$.

\section{Post-Schwarzschild expansion of EOB scattering}

If we now go back to the comparable mass case ($\nu \neq 0$), Eqs. \eqref{chiPM1+2}, \eqref{chi1PM} (with $\e$ defined by Eq. \eqref{f})
for the 1PM ($O(G)$) contribution to the scattering function,
display the main result of Ref. \cite{Damour:2016gwp}: namely, the 1PM real dynamics is  fully encoded (at order $O(G)$) in the following two EOB ingredients: (i) the energy map \eqref{f} between
the real energy  $\E$ (and the real Hamiltonian $H_{\rm real}$) of the two-body system, and the effective energy $\Ef$ (and the
effective Hamiltonian $H_{\rm eff}$); and (ii) the determination of the effective Hamiltonian $\Ef = H_{\rm eff}({\bf R}, {\bf P})$ from the
mass-shell condition satisfied by an effective particle of mass $\mu = m_1 m_2/(m_1+m_2)$, and conserved energy $\Ef = - P_0$ following a geodesic in 
a Schwarzschild metric of mass $M$. We can then parametrize the 2PM, and higher PM, corrections to the dynamics by considering
general deformations of the latter Schwarzschild-like mass-shell condition, i.e. a generalized mass-shell condition of the type
\be \label{massshellgen}
0= g_{\rm eff}^{\mu \nu} P_{\mu} P_{\nu} + \mu^2 + Q\,,
\ee
where $ g_{\rm eff}^{\mu \nu}$ is the (inverse of an) effective metric of the form
\be \label{geff}
 g^{\rm eff}_{\,\mu \nu} dx^{\mu} dx^{\nu} = - A dT^2 + B dR^2 + C (d \theta^2 + \sin^2 \theta d \varphi^2),
\ee
and where $Q$ is a Finsler-type additional contribution, which contains higher-than-quadratic in  momenta
contributions.

In previous EOB work, it has been standard to use deformed mass-shell conditions of the type \eqref{massshellgen}, involving
effective-metric functions $A,B,C$ that were $\nu-$deformed versions\footnote{Actually, it was found convenient to use a Schwarzschild-type
coordinate gauge where $C$ was fixed to $C_0=R^2$.} of the Schwarzschild metric functions $A_0,B_0,C_0$
entering \eqref{g0}, and to constrain the additional contribution $Q$ to be at least quartic in the momenta: $Q= O(\bP^4)$. 
In the present work, we find convenient to relax the constraint that  $Q$ be at least quartic in momenta,
and allow it to be a general even function of ${\bf P}$ (depending also on ${\bf R}$). Such a general $Q({\bf R}, {\bf P})$ can then absorb any quadratic-in-momenta, $\nu$-dependent
deformation which was previously attributed to the metric functions $A$ and $B$.  In the following, 
we shall allow $Q$ to start at order $O(\bP^2)$. We can then, without loss of generality, 
assume that the effective-metric functions $A,B,C$ actually coincide with the
Schwarzschild ones $A_0,B_0,C_0$. [To keep open the possibility of being more general, we shall, however, continue to denote
them simply as $A,B,C$.] 

The explicit form of the deformed mass-shell condition reads
\be \label{massshellgen2}
0=- \frac{\Ef^2}{A} + \frac{P_R^2}{B}+ \frac{P_{\varphi}^2}{C} + \mu^2 + Q \,.
\ee
Solving this mass-shell condition for $\Ef = - P_0$ then yields the effective Hamiltonian $H_{\rm eff}(\bP)=\Ef = - P_0$.
Namely, its square reads
\be\label{Hf2gen}
H_{\rm eff}^2({\mathbf R}, {\mathbf P})=A \left[ \mu^2 + \frac{P_R^2}{B}+ \frac{P_{\varphi}^2}{C} + Q  \right].
\ee
In view of the recent proof \cite{Damour:2016gwp} of the exactness of the energy map \eqref{f}, the corresponding
real Hamiltonian is
\be \label{Heob}
H_{\rm real}({\mathbf R}, {\mathbf P})=M \sqrt{1+2\nu \left(\frac{H_{\rm eff}}{\mu }-1 \right)}\,.
\ee

Finally, we parametrize here all the PM effects beyond the 1PM level by considering a general function $Q$ decreasing
at least like $1/R^2$ when $R \to \infty$. [Indeed, the $O(1/R)$ terms contained in the Schwarzschild metric functions
have been shown to fully describe the 1PM effects \cite{Damour:2016gwp}.]

Assuming that the deformed mass-shell condition \eqref{massshellgen}, i.e. \eqref{massshellgen2}, is solved for $P_R$ as
a function of $\Ef$ and $J= P_{\varphi}$, the scattering function $\chi(\Ef, J)$ is given by a formula precisely similar to the one used above
for Schwarzschild scattering, namely
\be
\pi + \chi(\Ef, J) = -  \frac{\partial }{\partial J}\int dR  \, P_R(R ; \Ef, J)  \,.
\ee

To use this exact, formal result, we need to approach it perturbatively. Instead of considering its straightforward PM
expansion (i.e. its expansion in powers of $G$), we shall consider what can be called its ``post-Schwarzschild" expansion.
In the mass-shell condition \eqref{massshellgen2} we consider the Schwarzschild functions $A, B, C$ as being exact 
(without expanding them in powers of $G$), but we treat $Q$ as a formally small quantity.

As it has been shown that, to linear order in $G$, the two-body scattering was described by an
effective metric equal to a Schwarzschild metric (of mass $M=m_1+m_2$), the post-Schwarzschild deformation
$Q$ starts at order $G^2$, and can therefore be written as a perturbative PM expansion of the type 
$Q \sim G^2 + G^3 + G^4 + \cdots$. Correspondingly, we can view the solution $P_R(\Ef, J)$ of the
mass-shell condition \eqref{massshellgen2} as having a perturbative expansion in powers of $Q = O(G^2)$ of the form
\bea
P_R(\Ef, J) &=& P_R^{(0)}(\Ef, J) + P_R^{(Q)}(\Ef, J) \nonumber \\
&& + \, P_R^{(Q^2)}(\Ef, J) + O(Q^3)\,.
\eea
Here $P_R^{(0)}(\Ef, J)$ is the solution of \eqref{massshellgen2} when $Q=0$, i.e.,
\be 
P_R^{(0)}(\Ef, J) = \pm \sqrt{\frac{B}{A}  } \sqrt{ \Ef^2 - A \left(\mu^2 + \frac{J^2}{C}\right) },
\ee
while $P_R^{(Q)}(\Ef, J)$ is the linear-in-$Q$ correction to the solution $P_R(\Ef, J)$ of \eqref{massshellgen2}, i.e.
\be
P_R^{(Q)}(\Ef, J) = - \frac{B}{2 P_R^{(0)}} Q \,.
\ee
As $Q$ starts formally at order  $G^2$, the contribution quadratic in $Q$ will be of order
\be
P_R^{(Q^2)}(\Ef, J) = O(Q^2) = O(G^4)\,.
\ee
The corresponding $Q$-expansion of the scattering function has the form
\bea
 \chi(\Ef, J) &=& \chi^{(0)}(\Ef, J) + \chi^{(Q)}(\Ef, J) \nonumber \\
 && + \,\chi^{(Q^2)}(\Ef, J) + O(Q^3)\,.
\eea
Here,
\be
\chi^{(0)}(\Ef, J) = -\pi   -  \frac{\partial }{\partial J}\int dR  \, P_R^{(0)}(R ; \Ef, J)  \,,
\ee
is simply the scattering function $\chi^{\rm Schw}(\Ef, J)$ in the Schwarzschild-type metric defined by the Schwarzschild-like functions $A,B,C$.
It is given by the formulas given in the previous Section, modulo the replacements 
\be
M_0 \to M \,; \, m_0 \to \mu \, ; \, {\mathcal E}_0 \to \Ef \, ; \, J_0 \to J\,.
\ee
We therefore conclude that the post-Schwarzschild contribution to $\chi(\Ef, J)$ is related to $Q$ via the simple formula
\be
\chi(\Ef, J) - \chi^{\rm Schw}(\Ef, J) = \chi^{(Q)}(\Ef, J) +  \chi^{(Q^2)}(\Ef, J) + \cdots
\ee
where
\be
\chi^{(Q)}(\Ef, J) =  - \frac{\partial }{\partial J}\int dR \, P_R^{(Q)}(\Ef, J) \,,
\ee
\be
\chi^{(Q^2)}(\Ef, J) =  - \frac{\partial }{\partial J}\int dR \, P_R^{(Q^2)}(\Ef, J) \,.
\ee
This yields
\be
\chi^{(Q)}(\Ef, J) =  + \frac{\partial }{\partial J}\int \frac{B dR}{2 P_R^{(0)}} Q\,,
\ee
and
\be
\chi^{(Q^2)}(\Ef, J) = O(Q^2) = O(G^4)\,.
\ee
The linear-in-$Q$ contribution can be rewritten as
\be
\chi^{(Q)}(\Ef, J) =  + \frac{\partial }{\partial J}\int \frac{ dR}{2 P^R_{(0)}} Q= + \frac{\partial }{\partial J}\int \frac12 d \sigma_{(0)} Q\,,
\ee
where $P^R_{(0)}$ now denotes the {\it contravariant} radial (unperturbed) momentum, i.e.\footnote{In the  expression 
below for $P^R_{(0)}$, the factor $1/\sqrt{AB}$ is actually equal to 1, but we did not use
this simplification to keep our formulas eventually applicable to a more general setting.
}
\be \label{PR0}
P^R_{(0)}= \frac1B  P_R^{(0)} = \pm \frac{1}{\sqrt{AB}} \sqrt{ \Ef^2 - A \left(\mu^2 + \frac{J^2}{C}\right) },
\ee
while $\sigma_{(0)}$ denotes the unperturbed (mass-normalized) effective propertime along the motion
\be
d \sigma_{(0)} = \frac{dR}{P^R_{(0)}} = \frac{ ds_{{\rm eff}  (0) } }{\mu}\,.
\ee
Indeed, along the unperturbed geodesic motion, we have $P^R_{(0)} = \mu \, dR/ds_{{\rm eff} (0) }$.

Combining the above results, and expressing them in terms of the scattering function 
$\frac12 \chi(\Ef, J)$ we have
\be
\frac12 \left( \chi(\Ef, J) - \chi^{\rm Schw}(\Ef, J) \right)= \frac14 \frac{\partial }{\partial J}\int  d \sigma_{(0)} Q + O(G^4) \,.
\ee
A first consequence of this result is that, within the accuracy indicated, the directly observable
scattering function only depends on the proper-time integral of the mass-shell perturbation $Q$.
In other words, modulo $O(G^4)$ the physics is invariant under transformations of the type
\be \label{Q'}
Q'({\bf R}, {\bf P})=Q({\bf R}, {\bf P})+ \frac{d}{d \sigma_{(0)}} G({\bf R}, {\bf P}) \,,
\ee
where the second term should be re-expressed in terms of  $({\bf R}, {\bf P})$ by using the
(at this order, unperturbed) equations of motion. It is easily seen that such a gauge-like transformation
of $Q$ corresponds to a (linearized) canonical transformation of the 
(Stueckelberg-like)  proper-time Hamiltonian
\be \label{calH}
{\cal H}(X^{\mu}, P_{\mu})= \frac12 \left( g_{\rm eff}^{\mu \nu}(X) P_{\mu} P_{\nu} + \mu^2 + Q(X,P) \right) \,.
\ee
Here, we allow $Q$ to depend on all components of $X^{\mu}$, and $P_{\mu}$. Above, we were 
generally assuming that $P_0$ was perturbatively replaced in terms of the spatial components 
${\bf P}$, so as to get more directly an ordinary Hamiltonian $ H_{\rm eff}({\bf R}, {\bf P})$ for the
evolution with respect to the effective time $T_{\rm eff}$.
We will use below the gauge freedom \eqref{Q'} to simplify the expression of $Q$.

The perturbative nature of the correlated PM expansions of $Q$ and $\chi$ is made clearer if we work 
 with the following dimensionless quantities (where the index $\mu$ on $p$ and $P$ should
be distinguished from the reduced mass $\mu = m_1 m_2/(m_1+m_2)$)
\be \label{rescaled}
p_{\mu} \equiv \frac{P_{\mu}}{\mu} \,;\, \widehat Q \equiv \frac{Q}{\mu^2}\, ; \, u \equiv \frac1r \equiv \frac{GM}{R} \,.
\ee
Using the basic fact that $u \equiv \frac{GM}{R}$ is of order $G$,  $\hQ$ will have a PM expansion the form
\be \label{hQPM}
\hQ= u^2 q_2(p) + u^3 q_3(p) + u^4 q_4(p) + \cdots \,,
\ee
where the term $ u^2 q_2(p) \propto G^2$ is of the 2PM level,  $u^3 q_3(p) \propto G^3$ of the 3PM level, etc.
For brevity, we have simply denoted  as $p$ the momentum-like arguments  that the various $q_n$'s depend upon.
Actually, $p$ could stand her for any (dimensionless, and time-symmetric)  scalar function of $p_{\mu}$ , ${\bf n} = {\bf R}/R$, and also $u$, that admits a finite limit as $u\to 0$. For instance, we could take a function of  $ ({\bf n} \times {\bf p})^2 = p_{\varphi}^2 u^2$ and $({\bf n} \cdot \p)^2 =p_r^2$,
but (as it will be integrated over the unperturbed scattering motion) we could also include a 
dependence on the energy $-p_0= \e$, considered along the unperturbed mass shell\footnote{We could also 
allow in $Q$ a dependence on $-p_0$ unrestricted by any mass-shell condition. This is, however, inequivalent (beyond the leading-order, $O(G^2)$, in PM perturbation theory) to using a dependence on $-p_0^{\rm on-shell}$.}, i.e. on the 
unperturbed effective Hamiltonian, 
\bea \label{HfS}
\e^{\rm on-shell}&=& {\widehat H}_{\rm eff}^{\rm Schw} \nonumber \\
&=& \!\! \sqrt{(1-2u) \left[1+(1-2u) ({\bf n} \cdot \p)^2 + ({\bf n} \times {\bf p})^2\right]} \nonumber \\
&=&\!\! \sqrt{(1-2u) \left[ 1 + (1-2u) p_r^2 + p_{\varphi}^2 u^2 \right] }.
\eea
Note in passing that we could also use a dependence on the unperturbed square kinetic energy 
$(1-2u) p_r^2 + p_{\varphi}^2 u^2$.

Transcribing the results above in terms of such dimensionless variables, we can relate the PM
expansion \eqref{chiPM1+2} of (half) the scattering function to the PM expansion \eqref{hQPM}
of $\hQ$ in the following way
\bea
&& \frac1{j^2} \left(\chi_{2}(\e, \nu) - \chi_{2}^{\rm Schw}(\e, \nu)\right) \nonumber \\
&& +  \frac1{j^3} 
 \left(\chi_{3}(\e, \nu) - \chi_{3}^{\rm Schw}(\e, \nu)\right) + O(\frac1{j^4})\nonumber\\
&& = \frac14 \frac{\partial }{\partial j}\int  
 \frac{dr}{p^r_{(0)}}  \hQ + O(G^4) \,.
\eea
Here, we used the fact that $\chi_{1}(\e, \nu) \equiv \chi_{1}^{\rm Schw}(\e, \nu)$. Note from
\eqref{hEj} that $1/j=O(G)$ so that the term $O(1/j^4)$ on the lhs is of the same order as the
$O(G^4)$ error term on the rhs (linked to the contribution quadratic in $Q$).

We shall explicitly check below that the integral $\int  \frac{dr}{p^r}  \hQ $ has a large-$j$
expansion of the type $\sim \frac1j + \frac1{j^2} = \cdots$. We can then integrate the above
result with respect to $j$ to get
\bea \label{chi23}
&& -\frac1{j} \left(\chi_{2}(\e, \nu) - \chi_{2}^{\rm Schw}(\e, \nu)\right) \nonumber \\
&& - \frac12  \frac1{j^2} 
 \left(\chi_{3}(\e, \nu) - \chi_{3}^{\rm Schw}(\e, \nu)\right) + O(\frac1{j^3})\nonumber\\
&& = \frac14 \int  \frac{dr}{p^r_{(0)}}  \hQ + O(\hQ^2) \,.
\eea

\section{Determining the effective Hamiltonian at the 2PM accuracy.}

Let us now focus on the contribution to the rhs of \eqref{chi23} brought by the 2PM term $u^2 q_2(p)$ in $\hQ$.
We recall  that the subscript $(0)$ added to $p^r$ indicates that (to linear order in $\hQ$) we
can neglect the effect of $\hQ$ in the integral $\int dr/p^r$, i.e. integrate over the Schwarzschild
scattering dynamics, with $p^r_{(0)}$ given by the following rescaled version of \eqref{PR0}
\bea \label{pr0}
p^r_{(0)}&=& \pm \frac{1}{\mu \sqrt{AB}} \sqrt{ \Ef^2 - A \left(\mu^2 + \frac{J^2}{C}\right) } \nonumber \\
&=& \pm  \sqrt{ \e^2 - (1-2u) (1 + j^2 u^2) }\,.
\eea
Inserting  $\hQ=u^2 q_2(p)$ on the rhs of \eqref{chi23} (and using $r=1/u$) yields
\be \label{intq2}
\frac14 \int  \pm \frac{du \,  q_2(p)}{\sqrt{ \e^2 - (1-2u) (1 + j^2 u^2) }}\,.
\ee
The integral here (as well as all integrals above) are to be taken over the full scattering motion,
with time going from $- \infty$ to $+\infty$, 
i.e over both the incoming motion (starting at $r= +\infty$ with $p^r<0$), and the outgoing one
(with $p^r>0$ back  to $r= +\infty$). We can simplify the evaluation of this integral by assuming
that we used a canonical transformation \eqref{Q'} such that the function 
$q_2(p)=q_2(({\bf n} \times {\bf p})^2, ({\bf n} \cdot {\bf p})^2; u)$ (considered along the
unperturbed, Schwarzschild mass-shell) depends only on the unperturbed effective Hamiltonian \eqref{HfS}, 
and is therefore constant during the integration over the scattering motion. At this order of approximation,
we could alternatively consider that $q_2(p)$ is only a function of, e.g., 
\be \label{p2vse}
{\bf p}^2=  p_r^2+ ({\bf n} \times {\bf p})^2 = \frac{\e^2}{1-2u}+ 2u p_r^2-1= \e^2 -1+ O(G)\,.
\ee
[However, if we were expressing $q_2(p)$ as a function of ${\bf p}^2$, 
the $O(G)$ correction in \eqref{p2vse} would modify the determination of the 3PM term $u^3 q_3(p)$.]

Assuming $q_2(p) = q_2(\e)$, we recognize on the rhs of \eqref{intq2} an integral giving
the scattering angle of a test particle in a Schwarzschild background. More precisely, we find (using
$u=y/j$ where $y$ was the integration variable used in Section \ref{secSchw})
\bea
\int_{- \infty}^{+ \infty}  \frac{dr}{p^r_{(0)}} u^2 &=& \int \frac{du}{\pm \sqrt{ \e^2 - (1-2u) (1 + j^2 u^2) }} \nonumber \\
&=& \frac1j \left[ \pi+ \chi^{\rm Schw}(\e,j) \right] \nonumber \\
&=&  \frac1j \left[ \pi+ 2 \frac{\chi_1^{\rm Schw}(\e) }{j} + O(\frac1{j^2})\right] ,
\eea
where we inserted the beginning of the PM expansion, derived in Sec. \ref{secSchw} above, 
of the Schwarzschild scattering angle. 

One  sees that we can neglect the fractional $O(1/j)$ correction linked to $\chi_1^{\rm Schw}(\e)$
when relating the post-Schwarzschild 2PM scattering angle  to $q_2(\e)$. We then get the very simple link
\bea \label{chi2}
&& -\frac1{j} \left(\chi_{2}(\e, \nu) - \chi_{2}^{\rm Schw}(\e)\right) \nonumber \\
&& = \frac14 \frac{\pi}{j}  q_2(\e)\,,
\eea
i.e.
\be \label{q2chi2}
q_2(\e , \nu)= - \frac{4}{\pi} \left[ \chi_{2}(\e, \nu) - \chi_{2}^{\rm Schw}(\e)  \right].
\ee
When considering the 3PM contribution to the scattering angle, and the corresponding 3PM
contribution to $\hQ$, expressed in terms of the unperturbed Hamiltonian $\e^{\rm on-shell}$,
i.e.
\be \label{hQPM}
\hQ = u^2 q_2( {\widehat H}_{\rm eff}^{\rm Schw}) + u^3 q_3( {\widehat H}_{\rm eff}^{\rm Schw})+ O(u^4),
\ee
we find the following link
\bea \label{q3chi3}
q_3(\e , \nu)&=&\frac{4}{\pi} \frac{2 \, \e^2-1}{\e^2-1} \left( \chi_{2}(\e, \nu) - \chi_{2}^{\rm Schw}(\e)   \right) \nonumber \\
&-& \frac{\chi_{3}(\e, \nu) - \chi_{3}^{\rm Schw}(\e)}{\sqrt{\e^2-1}} \,.
\eea
Note that, as the Schwarzschild scattering is the $\nu \to 0$ limit of the two-body one, the above expressions
for $q_2$ and $q_3$ explicitly show that 
\be
\lim_{\nu \to 0} q_2(\e , \nu)=0 = \lim_{\nu \to 0} q_3(\e , \nu) .
\ee
Summarizing: from Eq. \eqref{Hf2gen}, the squared effective Hamiltonian has the form
\be \label{hHf2bis}
{\widehat H}^2_{\rm eff}({\bf r}, {\bf p}) = {\widehat H}^2_{\rm Schw} + (1-2u) \hQ,
\ee
where
\be \label{hHS}
{\widehat H}^2_{\rm Schw}(p_r,r,p_\varphi ) \equiv (1-2u) \left[ 1 + (1-2u) p_r^2 + p_{\varphi}^2 u^2 \right] \,,
\ee
and where the PM-expansion of $\hQ$ is given by Eq. \eqref{hQPM}. In the latter PM-expanded value of $\hQ$, the explicit expression 
for the function $ q_2( {\widehat H}_{\rm eff}^{\rm Schw})$ reads (after inserting the 2PM scattering angle 
\eqref{chi2PM} in the link \eqref{q2chi2}) 
\be \label{q2fin}
q_2(\hHfS , \nu)= \frac{3}{2} \left( 5 \, \hHfS^2-1 \right) \left[ 1 -  \frac{1}{\sqrt{1+ 2\nu (\hHfS-1)} } \right]\,.
\ee

In other words, the PM expansion (or, in fact, the post-Schwarzschild expansion) of the 
dimensionless squared effective Hamiltonian can be written (after a suitable canonical transformation) as
\bea \label{Hf2PM}
&& {\widehat H}^2_{\rm eff}(p_r,r,p_\varphi ; \nu) = {\widehat H}^2_{\rm Schw} + \nonumber \\
 &&(1-2u) \left[  u^2  q_2({\widehat H}_{\rm Schw} , \nu)  \right. \nonumber \\
&& \left. + u^3 q_3({\widehat H}_{\rm Schw}, \nu) + u^4 q_4({\widehat H}_{\rm Schw}, \nu) +\cdots \right],
\eea
where $u\equiv 1/r$,  where $\hHfS$ is defined by Eq. \eqref{hHS}, and where the function $q_2$ is given
by Eq. \eqref{q2fin}, while the function $q_3$ is currently unknown, but is deducible from the 3PM
scattering function via Eq. \eqref{q3chi3}. Similarly $q_4(\e)$ would be deducible from the 4PM scattering function
(but one would have to take into account the nonlinear effects in $Q$ in the derivation above).

Note in \eqref{Hf2PM} the presence of the overall factor $1-2u$ in front of the $u^2$, $u^3$, $u^4$ terms 
coming from combining Eq. \eqref{hHf2bis} with the definition  of $\hQ=u^2 q_2(\e)+ u^3 q_3(\e) + \cdots$. Evidently,
when working at the 2PM accuracy, one could approximate, modulo the 3PM contribution $O(u^3)$,
the function $(1-2u)  u^2 q_2({\widehat H}_{\rm Schw} , \nu)$ entering the 2PM-accurate Hamiltonian 
simply by $ u^2 q_2({\widehat H}_{\rm Schw} , \nu)$.

\section{Comparing and contrasting the 2PM Hamiltonian to previous results}

Let us first compare the 2PM-accurate effective Hamiltonian \eqref{Hf2PM} to the corresponding PN-expanded
effective Hamiltonian. Here, we shall focus on the 3PN-accurate effective Hamiltonian \cite{Damour:2000we,Damour:2001bu}  (see \cite{Damour:2015isa} for the 4PN-accurate effective Hamiltonian). In order to compare the 
PM-expanded result \eqref{Hf2PM} to the corresponding 3PN-expanded Hamiltonian we need to apply a suitable canonical
transformation. Indeed, Ref. \cite{Damour:2000we} has used a gauge where the quartic-in-momenta terms
in the post-Schwarzschild contribution $\hQ({\bf x}, {\bf p})$ to the mass-shell condition were transformed so
as to involve only $p_r^4 = ({\bf n} \cdot {\bf p})^4$. This type of gauge is rather different from the one
we found convenient to use above. It is straightforward to construct a PN-expanded canonical transformation between
the two types of gauge; it is of the form
\be
g(r,p_r, p_\varphi)= p_r \left( \frac1{c^4} \frac{g_0}{r} + \frac1{c^6} \left[ \frac{g_1}{r^2} + \frac{g_2 \, p_\varphi^2}{r^3} + \frac{g_3 \, p_r^2}{r}\right] \right),
\ee
with, for instance, $g_0=\frac32 \nu$ at the 2PN level, and similar $O(\nu)$ coefficients $g_1, g_2, g_3$
at the 3PN level.

Using such a gauge transformation, the 3PN effective Hamiltonian of \cite{Damour:2000we} can be put 
in the form of \eqref{Hf2PM} with PN-expanded versions of the various coefficients $q_2, q_3$ and $q_4$.
[Indeed, the PN-expansion of the contribution $u^4 q_4({\widehat H}_{\rm Schw}, \nu)$ starts at
the 3PN order, while the next term $u^5 q_5({\widehat H}_{\rm Schw}, \nu)$ would start at the 4PN level.]
We then found that the PN-expanded version of $q_2({\widehat H}_{\rm Schw}, \nu)$ obtained from
the 3PN-accurate Hamiltonian was in full agreement with the PM-exact expression \eqref{q2fin}, while the
PN-expanded versions of the currently unknown next PM terms $ q_3$ and $q_4$ were given by
\bea
q_3^{\rm PN}({\widehat H}_{\rm Schw}, \nu) &=& 5 \nu + \frac14 (108 \nu - 23 \nu^2) (\hHfS^2-1) \nonumber \\
&+&  O\left( (\hHfS^2-1)^2 \right) \,,
\eea
\bea
q_4^{\rm PN}({\widehat H}_{\rm Schw}, \nu) &=&  \nu \left( \frac{175}{3} - \frac{41}{32} \pi^2 \right) - \frac72 \nu^2  \nonumber \\
&+&  O\left( \hHfS^2-1 \right) \,.
\eea
Here, we have used as PN expansion parameter 
\be
\hHfS^2-1 = O(u) + O({\bf p}^2) = O(\frac1{c^2}) \,.
\ee
Concerning the PN expansion of $q_2$ note that it starts as 
\be
q_2^{\rm PN}({\widehat H}_{\rm Schw}, \nu)= 6 \nu \left( {\widehat H}_{\rm Schw} -1 \right) + O\left( (\hHfS-1)^2 \right) \,,
\ee
where we now used 
\be
\hHfS -1 = \frac{\hHfS^2-1}{\hHfS+1} \sim \frac12 \left(\hHfS^2-1 \right) \,,
\ee
as PN expansion parameter.

An important information contained in our 2PM-accurate result \eqref{q2fin} for $q_2({\widehat H}_{\rm Schw}, \nu)$
is that, while its PN expansion leads to a $\nu$-expansion of the type $q_2(\nu) \sim \nu + \nu^2 + \nu^3 + \cdots$,
its exact PM form shows that this $\nu$ expansion is {\it non uniformly valid} in phase-space, and actually
breaks down at high energies. More precisely, when the product $\nu \hHfS$ becomes of order unity the $\nu$-dependence of $q_2$ changes character. Most importantly, when the effective energy tends to infinity we have
\be \label{q2HE}
\lim_{\hHfS \to \infty} q_2(\hHfS , \nu) \approx \frac{15}{2}  \hHfS^2 \,,
\ee
where the rhs becomes {\it independent} of $\nu$. As we shall see in the next Section, such a 
large-energy behavior applies also to the higher PM contributions $q_n(\hHfS)$, which are expected
to behave as
\be \label{qnHE}
\lim_{\hHfS \to \infty} q_n(\hHfS , \nu) \approx c^{(q)}_n \hHfS^2 \,,
\ee
with  purely numerical $\nu$-independent coefficients $ c^{(q)}_n$. 

\section{High-energy limit of two-body scattering and two-body dynamics.}
\label{sec:HE1}

Let us start by noting that the high-energy (HE) limit ($\e \to \infty$) of the two-body scattering function
evaluated in the effective 1PM-accurate metric (defined here by neglecting $Q=O(G^2)$ in Eq. \eqref{massshellgen}) ,
i.e. the HE limit of the scattering of a particle of mass $\mu= m_1m_2/(m_1+m_2)$ in a Schwarzschild metric 
of mass $M=m_1+m_2$, has the form (from  Section \ref{secSchw})
\bea \label{chiSHE}
&&\frac12 \chi^{Q\to 0}(E_{\rm real}, J ; m_1, m_2, G) \, \overset{\rm HE}{=} 2 \frac{\e}{j} + \frac{15 \pi}{8} \frac{\e^2}{j^2} \nonumber \\
&& + \frac{64}{3} \frac{\e^3}{j^3} + \frac{3465 \pi}{128} \frac{\e^4}{j^4} + O\left( \frac{\e^5}{j^5} \right) \,.
\eea
Here, and below, the HE limit means $\e \to \infty$, $j \to \infty$ with $\e/j$ fixed. The indication ``HE" above an equal sign indicates an equality holding in the HE limit.

When adding to this result the effect of the 2PM-accurate value of $Q$ (namely $\hQ= u^2 q_2(\hHfS)$ with \eqref{q2fin}), it takes the new form
\bea \label{chiq2HE}
&&\frac12 \chi^{Q^{2\rm PM} }(E_{\rm real}, J ; m_1, m_2, G) \, \overset{\rm HE}{=} 2 \frac{\e}{j} + 0 \frac{\e^2}{j^2} \nonumber \\
&& + c_3 \frac{\e^3}{j^3} + c_4 \frac{\e^4}{j^4} + O\left( \frac{\e^5}{j^5} \right) \,,
\eea
where the numerical coefficient of the $O\left( \frac{\e^2}{j^2} \right)$ has been reduced to zero (because
of the factor $1/\sqrt{1 + 2 \nu (\e-1)} \overset{\rm HE}{=} 0$ in Eq. \eqref{chi2PM}), and where the coefficients
$c_3, c_4$, etc. are numerical coefficients that can, in principle, be deduced from our results, and which differ from the ones
in Eq. \eqref{chiSHE}.

It is clear from Eqs. \eqref{chiSHE} and \eqref{chiq2HE} that, in the HE limit, the scattering function
$\frac12 \chi^{Q\to 0}(E_{\rm real}, J ; m_1, m_2, G)$ does not depend on all variables it could a priori depend,
but is only a function of the dimensionless ratio
\be \label{defalpha}
\alpha \equiv \frac{\e}{j} \equiv \frac{G M \Ef}{J}\,.
\ee
Using the EOB link between $\Ef$ and the real two-body energy $\E$, namely
\be \label{f'}
\Ef= \frac{({E}_{\rm real})^2 - m_1^2  -m_2^2 }{2 \, (m_1 + m_2)}\,,
\ee
we then see that we can re-express the expansion parameter $\alpha$ as
\be
\alpha = \frac{G}{2} \frac{({E}_{\rm real})^2 - m_1^2  -m_2^2 }{J} \overset{\rm HE}{=} \frac12 \frac{G ({E}_{\rm real})^2 }{J} \,.
\ee
Note that, in the HE limit, when $\alpha$ is expressed in terms of the real two-body c.m. energy,
and the real two-body c.m. angular momentum, it no longer depends on the masses, but only on the
dimensionless combination $ G \E^2/J$. Another useful expression for $\alpha$ consists in using the
c.m. impact parameter $b$, which is such that
\be \label{jbp}
J = b \, P_{\rm c.m.} \,.
\ee
The c.m. energy is the following function of the c.m. momentum\footnote{We use here an upper case $P$
as a reminder that $P_{\rm c.m.}$ is not rescaled by $\mu$ as the EOB momentum 
$\p = {\bf P}^{\rm EOB}/\mu$, Eq. \eqref{rescaled}.} $P_{\rm c.m.}$
\be \label{ecm}
\E = \sqrt{m_1^2+  P_{\rm c.m.}^2} + \sqrt{m_2^2+  P_{\rm c.m.}^2} \overset{\rm HE}{=} 2 \,  P_{\rm c.m.} \,,
\ee
so that we can also write the HE limit of $\alpha$ as
\be
\alpha \overset{\rm HE}{=} \frac{G \E}{b} \,.
\ee
The latter expression makes it particularly clear why, during a HE collision, the scattering angle should only
depend on $\alpha$. From the point of view of EOB theory, the important fact contained in the different expressions
above for $\alpha$ is that it shows the compatibility between an effective particle description where the total rest mass
$M$ plays an explicit role (namely $\alpha = G M \Ef/J$), and the standard way of looking at a HE collision where
one would instead  expect the mass-independent dimensionless parameter 
$ Gs/J \equiv G (\E)^2/J \overset{\rm HE}{=} 2 \alpha$ to be the controlling parameter. Note that the compatibility
between the two descriptions crucially relies on the quadratic nature of the EOB energy map \eqref{f'}. This is a new
(HE) check of the fact that this energy map is exact.

It is easy to see that the structure of  general HE expansions of the type of Eqs. \eqref{chiSHE} or \eqref{chiq2HE} 
is a direct consequence of having a mass-shell condition that is {\it quadratic} in the momenta in the HE limit. Indeed,
if we take neglect the rest-mass term $m_0^2$ in the equations of Section \ref{secSchw}, and rewrite the results there 
for arbitrary metric functions\footnote{and replacing ${\mathcal E}_0 \to \Ef$, $J_0 \to J$.}  $A(R), B(R), C(R)$ 
(in lieu of only $A_0(R), B_0(R), C_0(R)$), it is clear from the start that only the ratio $\Ef/J$ matters. More explicitly,
in the HE limit, the final formula \eqref{chigen1} reads 
\bea \label{chigen1HE}
\pi &+& \chi \overset{\rm HE}{=}  \int J\frac{dR}{C} \frac{\sqrt{A B} }{\pm \sqrt{{\mathcal E}_{\rm eff}^2 -J^2 \frac{A}{C}  } } \nonumber \\
&& = \int \frac{dR}{C} \frac{\sqrt{A B} }{\pm \sqrt{\frac{{\mathcal E}_{\rm eff}^2}{J^2} -\frac{A}{C}  } }\,.
\eea
In addition, if we use, for simplicity, a coordinate gauge where $C(R)=R$, and if the coefficients $A(R), B(R)$ of the 
effective metric depend on $R$ only through the dimensionless
combination $u=GM/R$ (involving $G M=G(m_1+m_2)$ as length scale), we can rewrite \eqref{chigen1HE} as
\be \label{chigen2HE}
\frac{\pi}{2}+\frac{\chi}{2} \overset{\rm HE}{=} \int_0^{u_{\rm max}(\alpha)} du \frac{\sqrt{A(u) B(u)} }{\pm \sqrt{\alpha^2 - u^2 A(u) } }\,,
\ee
where $\alpha$ is defined by Eq. \eqref{defalpha}, and where we now restricted the integration range
to the interval $0<u<u_{\rm max}(\alpha)$, where $u_{\rm max}(\alpha)$ is the (positive) root of 
$u^2A(u) = \alpha^2$ closest to the origin. This shows explicitly that, in the HE limit, $\chi$ depends only
on $\alpha$.

We have seen above that, in the HE limit, the rather involved momentum dependence of the 2PM-accurate mass-shell
condition (which involves the complicated function \eqref{q2fin}) drastically simplified.  More precisely,
inserting the HE limit \eqref{q2HE} (and its higher PM analog \eqref{qnHE}) in the mass-shell condition \eqref{massshellgen} (in which we recall
that $A,B,C$ denote the Schwarzschild metric functions), and neglecting the rest-mass contributions,
 we get the following simple HE mass-shell condition $\e^2$
\bea \label{HEmassshell}
0 &=& - \frac{\Ef^2}{1-2u}+ K_{\rm Schw} \nonumber \\
&+& \left(\frac{15}{2} u^2 + c_3^{(q)} u^3 + \cdots \right)  (1-2u) K_{\rm Schw},
\eea
where we denoted the Schwarzschild-like kinetic-energy by
\be
K_{\rm Schw} \equiv (1-2u) P_R^2 +\frac{P_\varphi^2}{R^2}\,.
\ee
The HE mass-shell condition \eqref{HEmassshell} is {\it quadratic} in momenta. If we introduce the function
\be \label{fvsq}
f(u) \equiv (1-2u) \left(\frac{15}{2} u^2 + c_3^{(q)} u^3 +  c_4^{(q)} u^4 +\cdots \right),
\ee
the HE mass-shell condition reads
\be
0 = - \frac{\Ef^2}{1-2u}+ (1+ f(u)) K_{\rm Schw}\,,
\ee
or, equivalently,
\be
0= - \frac{\Ef^2}{A_{\rm HE}(u)}+ K_{\rm Schw}=  - \frac{\Ef^2}{A_{\rm HE}(u)}+ (1-2u) P_R^2 +\frac{P_\varphi^2}{R^2},
\ee
where we further defined
\be \label{AHE}
A_{\rm HE}(u)\equiv (1-2u) \left[ 1+f(u) \right] \,.
\ee
In other words, we see that the HE limit of the scattering is equivalent to a {\it null geodesic} in the
``effective HE metric"
\be \label{geffHE}
ds^2=- A_{\rm HE}(u) d T^2 + \frac{dR^2}{1-2u} + R^2 (d \theta^2 + \sin^2 \theta d \varphi^2),
\ee
which differs from the Schwarzschild metric only through the deformed time-time coefficient $A_{\rm HE}(u)$,
given by Eq. \eqref{AHE}.

Our 2PM calculations above have only given us access to the $O(u^2)$ contribution to 
the correcting factor $1+ f(u) =A_{\rm HE}(u)/A_{\rm Schw}(u)$ to $A_{\rm Schw}(u)=1-2u$,
namely
\bea \label{f2PM1}
  1+ f^{\rm 2PM}(u)&=&  1+(1-2u)\frac{15}{2} u^2 + O(u^3)   \nonumber \\
&=&   1+\frac{15}{2} u^2 + O(u^3)  \,.
\eea

Let us now show how to derive a more accurate value of $f(u)$ from the ultrahigh-energy scattering
results of Amati, Ciafaloni and Veneziano \cite{Amati:1990xe}. Indeed, Ref. \cite{Amati:1990xe} 
evaluated at two loops (using  an eikonal expansion)  the HE scattering angle scattering of two gravitons
(or low-mass string states)\footnote{As the HE scattering is blind to the rest masses, one can consider
the scattering of gravitationally interacting massless particles such as gravitons.}. In terms of our parameter $\alpha$, Eq. \eqref{defalpha}, their result [Eq. (5.28) in \cite{Amati:1990xe}] reads
\be
\sin \frac12\chi^{\rm ACV}  \overset{\rm HE}{=} 2 \alpha + (2 \alpha)^3 + O(\alpha^5) \,,
\ee
or, equivalently
\be \label{chiACV}
\frac12\chi^{\rm ACV}  \overset{\rm HE}{=} 2 \alpha + \frac76 (2 \alpha)^3 + O(\alpha^5) \,.
\ee
Note the information given by Amati, Ciafaloni and Veneziano that, because of the analyticity properties in 
$s=\E^2$ of scattering amplitudes, there are no contributions of order $\alpha^4$. 

We need to compare Eq. \eqref{chiACV} to the lightlike scattering in the effective metric \eqref{geffHE}, i.e. (from \eqref{chigen2HE})
\be \label{chigen2HE}
\frac{\pi}{2}+\frac{\chi}{2} \overset{\rm HE}{=} \int_0^{u_{\rm max}(\alpha)} du \frac{\sqrt{A_{\rm HE}(u) B(u)} }{ \sqrt{\alpha^2 - u^2 A_{\rm HE}(u) } },
\ee
where $A_{\rm HE}(u)$ is given by Eq. \eqref{AHE}. Parametrizing the PM expansion of $f(u)$ as
\be
f(u) = f_2 u^2 + f_3 u^3 + f_4 u^4 + \cdots \,
\ee
one can compute the integral in Eq. \eqref{chigen2HE} in terms of the numerical coefficients $f_n$ and compare
the result to \eqref{chiACV}. A convenient way of computing the integral \eqref{chigen2HE} is to replace
the integration over the variable $u$ by an integration over the variable $x$ defined so that
\be
\alpha^2 x^2 = u^2 A_{\rm HE}(u) \,.
\ee
This reduces the evaluation of the integral in Eq. \eqref{chigen2HE} to an integral of the type
\be
\int_0^1 \frac{dx}{\sqrt{1-x^2}} \left( 1+ c_1 \alpha x + c_2 (\alpha x)^2 + \cdots \right)\,,
\ee
where the coefficients $c_n$ are linear combinations of the $f_n$'s. One finally deduces that the
result of Amati, Ciafaloni and Veneziano implies the following PM expansion of the correction factor $f(u)$:
\be \label{fACV}
f^{\rm ACV}(u) = \frac{15}{2} u^2 - 18 u^3 + \frac{1845}{16} u^4 + O(u^5) \,.
\ee
In other words, using the link \eqref{fvsq}, this implies that the HE limit of the function $(1-2u) \hQ$
is of the form
\be
(1-2u) \hQ \overset{\rm HE}{=} \left(  \frac{15}{2} u^2 - 18 u^3 + \frac{1845}{16} u^4 + O(u^5)\right) \hHfS^2 \,.
\ee
This is equivalent to
\be
\hQ \overset{\rm HE}{=} \left(  \frac{15}{2} u^2 - 3 u^3 + \frac{1749}{16} u^4 + O(u^5)\right) \hHfS^2 \,.
\ee 
The agreement between the 2PM contribution ($\frac{15}{2} u^2$) in the results so derived from
the 2-loop computation of Ref. \cite{Amati:1990xe} is an independent check of our results above.
However, this is only a check of the vanishing of the $\alpha^2$ contribution to the
HE limit of $\chi$ (due to the $1/\sqrt{1+ 2\nu (\e-1)}$ factor in  Eq. \eqref{chi2PM}).
It is remarkable that the results of Ref. \cite{Amati:1990xe} allow one to derive non trivial information
about the HE behavior of $Q$ at the 3PM and 4PM levels.

\section{Self-force expansion and  light-ring behavior.}

The EOB formalism was initiated by starting from the PN-expanded dynamics, with the aim
of extending its validity beyond the range of applicability of PN theory (slow velocities and small gravitational
potentials) so as to describe the last orbits and the coalescence of binary black holes. One of the first results of
EOB theory was to find that, though the end of the inspiral motion is non-adiabatic and involves a non-negligible
radial motion of the coalescing bodies, the kinetic energy associated with this radial motion remains
rather small compared to the kinetic energy of the angular motion, {\it even during the `plunge' phase}
which follows the crossing of the last stable (circular) orbit (LSO) \cite{Buonanno:2000ef}. This fact motivated
Damour, Jaranowski and Sch\"afer (DJS), when they found that the 3PN extension of the EOB dynamics necessitated
the introduction of quartic-in-momenta contributions to the effective mass shell (i.e. a term $Q = O({ P}^4)$ 
in Eq. \eqref{massshellgen}), to use a canonical transformation reducing the $P^4$ dependence of $Q$ 
(which would a priori involve ${\bf P}^4$, $({\bf n} \cdot {\bf P})^2  {\bf P}^2$, and $({\bf n} \cdot {\bf P})^4$)
to a dependence on the sole radial kinetic energy term, i.e. $ P_R^4 \equiv ({\bf n} \cdot {\bf P})^4$ \cite{Damour:2000we}.
This ``DJS gauge" was recently extended to the 4PN level \cite{Damour:2015isa}. It was shown in Ref. \cite{Damour:2000we}, by using a counting argument, that there formally existed, at all PN orders, a PN-expanded canonical 
transformation able to reduce the $P$ dependence of $Q$ to a dependence on the sole $P_R$. 

The use of such a DJS gauge allowed EOB theory to pack the description of the energetics of circular orbits
into  the single EOB radial function ${\bar A}( \bu; \nu) \equiv - g^{\rm eff}_{00}({\bar R})$, where $\bu\equiv GM/(c^2 {\bar R})$. [Here, we added a bar both over $A$ and over the usual EOB gravitational potential to 
distinguish the value of this radial potential in the DJS gauge
(denoted ${\bar A}( \bu)$) from its value in the energy gauge (denoted simply $A(u)$) that we use in this paper when discussing 
the 2PM EOB dynamics.]
This description turned out to be quite convenient for finding good resummations of the PN expansion
\be
{\bar A}^{\rm PN}(\bu;\nu)= 1 - 2\bu + 2 \nu \, \bu^3 + a_4(\nu)  \,\bu^4 + a_5(\nu, \ln \bu)\, \bu^5 +\cdots
\ee
of the radial potential ${\bar A}( \bu; \nu)$. It also led to the discovery of remarkable cancellations leading to a 
dependence of ${\bar A}^{\rm PN}(\bu;\nu)$ on $\nu$ which is {\it linear} at the 2PN and 3PN levels, and no more than
quadratic at the 4PN level (while all the other functions describing the energetics of circular orbits involve
higher powers of $\nu$).[See a detailed discussion of this point in Ref. \cite{Bini:2013rfa}.]

More recently, the EOB formalism was directly extended into the strong-field regime by incorporating results
from self-force (SF) theory \cite{Damour:2009sm,Barack:2010ny,Barausse:2011dq,Akcay:2012ea,Bini:2014ica,Bini:2015xua,Hopper:2015icj,Kavanagh:2017wot}. 
Within the EOB framework, SF theory corresponds to expanding the various
EOB potentials (${\bar A}( \bu)$, ${\bar B}( \bu)$, ${\bar Q}( \bu, p_r)$) in a power series in $\nu$, e.g.:
\be
{\bar A}^{\rm SF}(\bu;\nu)= 1 - 2\bu +  \nu a_{1SF}(\bu) +  \nu^2 a_{2SF}(\bu) + O(\nu^3).
\ee
Current SF theory only allows one to compute the contributions linear in $\nu$, such as $a_{1SF}(\bu)$, but it
can (numerically) compute it even in the strong-field domain, i.e. for values of $\bu$ going even beyond the
LSO, up to the lightring (LR), i.e. $\bu_{\rm LR}= \frac13$, when considering the dynamics of a small mass
around a nonspinning black hole. [In that case, we recall
that ${\bar R}_{\rm LSO}= 6 GM/c^2 \to \bu_{\rm LSO}= \frac16$, while ${\bar R}_{\rm LR}= 3 GM/c^2 \to \bu_{\rm LSO}= \frac13$.] The first computation, at the first self-force (1SF) level, of a combination of EOB potentials
in the strong-field domain was achieved in Ref. \cite{Barack:2010ny}, and covered the interval $0< \bu < \frac16$, i.e. from
large values of ${\bar R}$ down to the LSO. The discovery of nice identities connecting the binary dynamics
to SF quantities \cite{LeTiec:2011ab,LeTiec:2011dp,Blanchet:2012at,Tiec:2015cxa} then allowed one to separately compute \cite{Barausse:2011dq} the 1SF contribution $a_{1SF}(\bu)$ to
the EOB $A$ potential in the interval $0< \bu < \frac15$. The (numerical) computation of $a_{1SF}(\bu)$
was later extended up to the LR, i.e. in the interval $0< \bu < \frac13$ \cite{Akcay:2012ea}, which is the largest interval
where SF theory can compute $a_{1SF}(\bu)$ (because this is the largest interval in which there exist 
circular orbits around a nonspinning black hole). [We recall that there exist {\it stable} circular orbits when
${\bar R}_{\rm LSO} < {\bar R} < + \infty$, and {\it unstable} circular orbits when ${\bar R}_{\rm LR} < {\bar R} < {\bar R}_{\rm LSO}$.]

A surprising finding of Ref. \cite{Akcay:2012ea} was that  the 1SF contribution $a_{1SF}(\bu)$ to the EOB $A$ potential
had a {\it divergent} behavior at the LR\footnote{Ref. \cite{LeTiec:2011dp} had earlier suggested 
(from the extrapolation of a rational fit to numerical SF data in the interval $0<x<\frac15$) the existence of 
a singularity, at the LR, in the 1SF contribution to the redshift function $z_1(x; \nu)$ . However, as $z_1^{0SF}(x) = \sqrt{1-3x}$, the presence of a singularity of the type
$\partial z_1^{0SF}(x)/\partial x\sim (1-3x)^{-1/2}$ was naturally expected, and would have followed from
EOB theory with a LR-regular radial potential $A(u;\nu)$. The surprising fact is that the LR singularity
of $z_1^{0SF}(x)$ is stronger than expected by an extra factor $(1-3x)^{-1/2}$, i.e. of the type $z_1^{0SF}(x) \sim (1-3x)^{-1}$. }, namely
\be \label{a1sf}
a_{1SF}(\bu)  \underset{\bu \to \frac13}{\sim} \frac14 \zeta (1- 3 \bu)^{-1/2}, \quad {\rm with} \quad \zeta \approx 1 \,.
\ee
Ref. \cite{Akcay:2012ea} understood the origin of this divergence as coming from the divergent-energy behavior
of the small particle as it approaches the LR, and argued that not only was the energy-renormalized function
\be
a^{1SF}_E(\bu) \equiv \frac{a_{1SF}(\bu)}{{\widehat E}_S(\bu)},  \; {\rm where} \;{\widehat E}_S(\bu) = \frac{1-2\bu}{\sqrt{1-3\bu}}\,,
\ee
finite at the LR (namely $a^{1SF}_E(\bu) \to \frac34 \zeta$ as $\bu \to \frac13^-$), but that it seemed to be naturally, and smoothly, extendable beyond the LR (i.e. for $\bu > \frac13$),
and even beyond the horizon (located at $\bu=\frac12$). Moreover, it seemed probable that the natural extension of the
regularized function $a^{1SF}_E(\bu)$ would linearly vanish at the horizon (i.e. contain a factor $1-2\bu$).

In addition, Ref. \cite{Akcay:2012ea} showed that the singular behavior \eqref{a1sf} of $a^{1SF}_E(\bu)$ was
just a ``coordinate singularity in the EOB phase space" which  ``can be avoided by a suitable phase-space 
transformation that replaces it with an alternative regular description". The latter alternative, regular description,
suggested in Ref. \cite{Akcay:2012ea}, consists in abandoning the use of the (restricted) DJS-gauge, and in using
instead a gauge such that
\be
{\widehat H}^2_{\rm eff}(p_r,r,p_\varphi ; \nu) = {\widehat H}^2_{\rm Schw} + \nu [{\widehat H}^2_{\rm eff}]_{1SF}+ O(\nu^2),
\ee
with a post-Schwarzschild $Q$-type contribution ($(1-2u) \hQ= \nu [{\widehat H}^2_{\rm eff}]_{1SF}+ O(\nu^2)
$) of the type (see Eq. (139) in \cite{Akcay:2012ea})
\be  \label{Hf21sfnum}
\nu  [{\widehat H}^2_{\rm eff}]_{1SF}=\nu a^{1SF}_E(u)\frac{\hHfS^3}{1-2u} + O(\nu^2),
\ee
where $\hHfS$ is the Schwarzschild effective Hamiltonian (see Eq. \eqref{hHS}). The crucial feature of 
Eq. \eqref{Hf21sfnum} is the {\it cubic} dependence on $\hHfS$.

Let us compare the latter suggestion to our PM-expanded result above \eqref{Hf2PM}. If we perform the SF expansion
of our PM-expanded Hamiltonian \eqref{Hf2PM} (restricted to the 2PM contribution $q_2$, which is the only 
one currently known for arbitrary large velocities) we get
\bea
{\widehat H}^2_{\rm eff}&=& {\widehat H}^2_{\rm Schw} \nonumber \\
&+& \frac32 \nu (1-2u) u^2\left( 5 \, \hHfS^2-1 \right) \left( \hHfS-1 \right) \nonumber \\
&\times& \left[1 - \frac32 \nu (\hHfS-1) +  \frac52 \nu^2  (\hHfS-1)^2 + \cdots \right] .\nonumber \\
\eea
The two important points concerning this expansion are: (i) we  recover what was the main point suggested
in \cite{Akcay:2012ea}, namely that the singularity \eqref{a1sf} appearing, at the LR, $\bu=\frac13$,
in the DJS gauge can be transformed by using a different phase-space gauge into a large-energy behavior
of the 1SF Hamiltonian of the form $[{\widehat H}^2_{\rm eff}]_{1SF} \sim \nu \hHfS^3$; and (ii) we get the new information
that this bad HE behavior is tamed by higher SF contributions, as exemplified by the factor
\bea \label{SFexp}
1 &-&  \frac{1}{\sqrt{1+ 2\nu (\hHfS-1)}}  = \nu (\hHfS-1) \nonumber \\
 &-& \frac32 \nu^2 (\hHfS-1)^2 
+  \frac52 \nu^3  (\hHfS-1)^3 + \cdots
\eea
Note indeed that while the left-hand side  (lhs) is well behaved and actually tends to unity in the HE limit $\hHfS \to \infty$,
each term on the rhs is divergent in the HE limit. This confirms one of the points of Ref. \cite{Akcay:2012ea},
namely that the SF expansion is not an expansion in the sole  parameter $\nu$, but rather an
expansion in an energy-corrected version of $\nu$.  In the present 2PM case,
we explicitly see that we have an expansion in powers of $\nu (\hHfS-1) $. In view of the
results given above for the PN expansions of the higher-PM analogs of $u^2 q_2(\hHfS)$, we
more generally expect that a general PM term will have an SF expansion controlled by the parameter
\be \label{tildenu}
\tilde \nu \equiv \nu \hHfS \,.
\ee
[The main difference between $\nu (\hHfS-1) $ and $\nu \hHfS$ is that, a low energies, i.e. when doing
a PN expansion, $\nu (\hHfS-1) $ starts at order $O(\frac{\nu}{c^2}) $, while $\nu \hHfS= \nu + O(\frac{\nu}{c^2}) $.]

At the current stage, one can analytically control only the leading-order contribution to the coefficient of 
$\hHfS^3$ in the large-energy limit of the 1SF expansion of $\hHf^2$, namely
\bea \label{Hf21sf2PM}
\nu [{\widehat H}^2_{\rm eff}]_{1SF}^{2PM} &\overset{\rm HE}{=}&  \nu \left[ \frac{15}{2} u^2(1-2u) + O(u^3)\right] \hHfS^3 \nonumber\\
&\overset{\rm HE}{=}&  \nu \left[ \frac{15}{2} u^2 + O(u^3)\right] \hHfS^3 \,.
\eea
The numerical value at the LR, i.e. for $u=\frac13$ (corresponding to the HE limit for circular orbits), of the coefficient of $\nu \hHfS^3$ is, when using the second line of Eq. \eqref{Hf21sf2PM}, $\frac{15}{18} \approx 0.83333$. 
By comparison, the numerical SF computation of \cite{Akcay:2012ea} leads to a numerical coefficient at the LR
equal to (see Eq. \eqref{Hf21sfnum}) $ \left[ a^{1SF}_E(u)/(1-2u)\right]_{u=\frac13}= \frac94  \, \zeta \approx 2.25$. One should not expect (in absence of higher PM contributions) any close numerical agreement, but
it is satisfactory to find that the sign and the order of magnitude of the lowest-order\footnote{If we had kept the 
3PM-level correction $-2 u^3$ present in the first line of Eq. \eqref{Hf21sf2PM}, one would have obtained a result
smaller by a factor $3$.} PM contribution is consistent with the SF result.

\section{High-energy Regge behavior of the EOB Hamiltonian, numerical simulations and light-ring behavior.}

An interesting aspect of our result \eqref{Hf2PM} is that it opens the possibility of exploring the gravitational
interaction in the HE limit.  The derivation of Eq. \eqref{Hf2PM} assumed a situation of small-angle scattering,
but once we have transcribed this information in terms of the Hamiltonian \eqref{Hf2PM}, we can also discuss
a situation where two compact objects (say two black holes) orbit each other, say on circular orbits\footnote{We recall that we are considering here the conservative dynamics of a two-body system. We shall comment below on how
to use numerical simulations of the dissipative dynamics to gain information about the conservative interaction.}, 
at very high kinetic energies (corresponding, in the $\nu \ll 1$ case, to motion near the the LR). In  the
previous Section we considered the HE limit of the SF-expanded dynamics, i.e. we {\it first} expanded the Hamiltonian
\eqref{Hf2PM} in powers of $\nu$, and {\it then} took the HE limit. Here, we shall instead consider the HE limit
of the non-SF-expanded Hamiltonian \eqref{Hf2PM}. As already noticed around Eq. \eqref{SFexp}, these two limits
(HE and SF) {\it do not commute}, essentially because the Hamiltonian crucially involve $\nu$ in the form of
the energy-dependent combination \eqref{tildenu}.

Let us consider the energetics of the sequence of two-body circular orbits defined by the 2PM-accurate Hamiltonian
\eqref{Hf2PM} (i.e., keeping only $q_2$, and setting to zero the higher PM contributions $q_3, q_4$ etc.). The energetics
of the sequence of circular orbits can be encoded in various gauge-invariant functions. The conceptually simplest one
is the $E$-$J$ curve, i.e. the functional link between the total orbital angular momentum $J$ of the binary system,
and the total energy $\E$. [We recall in passing that several works have shown how to extract this gauge-invariant
curve from numerical simulations of both binary black holes and binary neutron stars, and have (successfully) compared it
to its usual EOB description \cite{Damour:2011fu,Bernuzzi:2014owa,Nagar:2015xqa}] As is well-known, the Regge approach to particle physics has shown the importance
of considering the squared total energy, i.e. Mandesltam variable $s\equiv \E^2$, as a function of $J$. As the EOB
energy map \eqref{f} essentially identifies (modulo an additive constant and some rescalings) $s$ to the effective energy
$\Ef= \mu \, \e$, we shall focus our attention here on the functional link between $\Ef$ and $J$, or, in rescaled
variables, between $\e = \Ef/\mu$ and $j = J/(GM\mu)= J/(Gm_1m_2)$. Let us immediately note that the Regge slope
$ds/dJ$ is given in terms of rescaled variables by
\be \label{reggeslope}
\frac{ds}{dJ}= \frac{d \E^2}{dJ}= \frac2G \frac{d\e}{dj}\,.
\ee
Therefore, modulo the simple (mass-independent) factor $2/G$, the slope of the dimensionless curve $\e(j)$ gives the
Regge slope $ds/dJ$.

\begin{figure}
\includegraphics[scale=0.7]{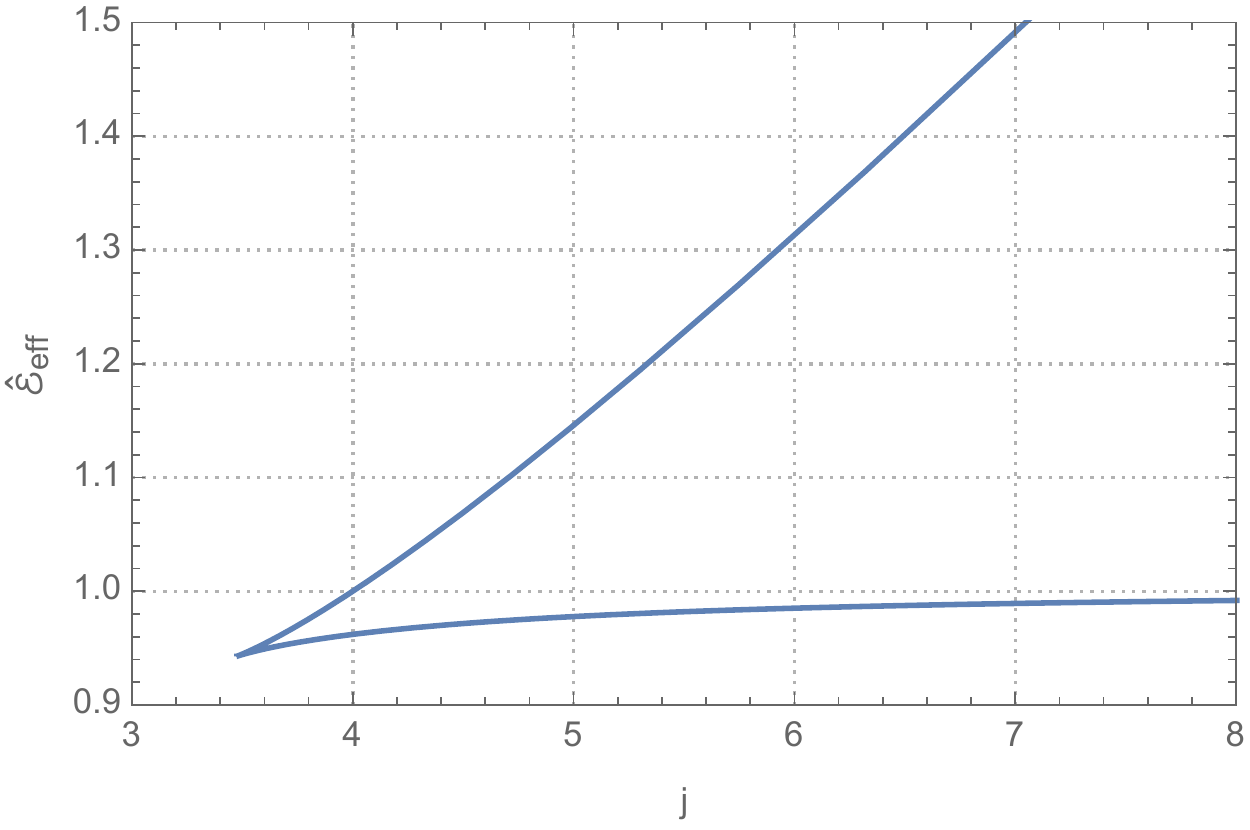}
\caption{\label{figej}
Graph of the relation between the rescaled angular momentum $j$ and the rescaled effective energy
$\hHf$ for $\nu=0.25$, and the 2PM Hamiltonian.}
\end{figure}

To get the 2PM-accurate functional link (along circular orbits) between $\e$ and $j$, one must eliminate the radial variable $u=1/r$
between the circular Hamiltonian $\hHf^{\rm circ}(u,j;\nu) \equiv \hHf(u, p_r=0,j;\nu)$, i.e.
\bea \label{Hf2PMcirc}
&& {\widehat H}_{\rm eff}^{2 \,\rm circ}(u,j ; \nu) = {\widehat H}^{2 \,\rm circ}_{\rm Schw} + \frac32 u^2(1-2u)\nonumber \\
 && \times \left( 5 \, {\widehat H}^{2 \,\rm circ}_{\rm Schw}-1 \right) \left[ 1 -  \frac{1}{\sqrt{1+ 2\nu ({\widehat H}^{\rm circ}_{\rm Schw}-1)} } \right],
\eea
where
\be \label{hHScirc}
{\widehat H}^{2 \,\rm circ}_{\rm Schw}(u,j) \equiv (1-2u) (1+ j^2 u^2) \,,
\ee
and the condition defining circular orbits, i.e.
\be
\frac{\partial {\widehat H}_{\rm eff}^{2 \,\rm circ}(u,j ; \nu)}{\partial u} =0.
\ee
We show in Fig. \ref{figej} the numerically computed $\e(j)$ curve, in the equal-mass case $\nu=0.25$. 
This curve is actually made of two branches: the lower
branch (such that $\e(j) \to 1^-$ as $j\to +\infty$) corresponds to the sequence of stable circular orbits (local 
{\it minima} of ${\widehat H}_{\rm eff}^{2 \,\rm circ}(u,j)$ for a fixed $j$); while
the upper branch (along which $\e(j) \to  +\infty$ as $j\to +\infty$) corresponds to the sequence of unstable
circular orbits (local 
{\it maxima} of ${\widehat H}_{\rm eff}^{2 \,\rm circ}(u,j)$ for a fixed $j$).
These two branches meet at a cusp which corresponds to the LSO. The location of the LSO for the case $\nu=\frac14$
is not very different from its test-mass limit $\nu \to 0$. Indeed, when $\nu \to 0$ we have the well-known
Schwarzschild value $u_{\rm LSO}(\nu=0) =\frac16=0.1666666$ corresponding to $j_{\rm LSO}(\nu=0) =\sqrt{12}=3.464102$ and $\e^{\rm LSO}(\nu=0) = \frac{2 \sqrt{2}}{3}=0.942809$. By contrast, when $\nu=\frac14$, we found 
$u_{\rm LSO}(\nu=\frac14) \approx 0.1666838$, corresponding to 
$j_{\rm LSO}(\nu=\frac14) \approx 3.474742$ and $\e^{\rm LSO}(\nu=\frac14) \approx 0.9428009$.
[The latter results are actually closer to their $\nu \to 0$ analogs than the ones following from the 2PN-expanded
EOB Hamiltonian \cite{Buonanno:1998gg}.]

 Our interest here is not in such quantitative results (which would be strongly modified by higher PM terms), 
 but rather in the new qualitative properties of the 2PM Hamiltonian (to be considered next)
 which follow from the large-energy behavior of the Hamiltonian \eqref{Hf2PMcirc}, and that are likely to
 hold also at higher PM orders.
 
 The first such qualitative result is the $\nu$-independence of the Regge slope  \eqref{reggeslope} in the HE limit.
 Indeed, when taking the  limit $j \to +\infty$ the $\nu$-dependent last (inverse squareroot) term in the circular Hamiltonian \eqref{Hf2PMcirc} tends to zero, so that we have the HE limit
 \be \label{Hf2circHE}
 {\widehat H}_{\rm eff}^{2 \,\rm circ}(u,j ; \nu) \overset{\rm HE}{=} j^2 u^2 (1-2u) \left( 1+ f^{\rm 2PM}(u) \right),
 \ee
 where, consistently with Eq. \eqref{f2PM1},
 \be \label{f2PM2}
  f^{\rm 2PM}(u)= \frac{15}{2} u^2 (1-2u) .
 \ee
 It is easily seen that maximizing the HE circular Hamiltonian \eqref{Hf2circHE} with respect to $u$ leads to $u=\frac13$
 (i.e. the $u$-location of the HE, 2PM-accurate, LR happens to be equal to its $\nu\to 0$ value). We then find that the HE limit
 of the (rescaled) Regge slope \eqref{reggeslope} is equal to
 \be \label{slope2PM}
\frac{d\e}{dj} \overset{\rm HE}{=} \left[\sqrt{ u^2 (1-2u) \left( 1+ f^{\rm 2PM}(u) \right) } \right]_{\rm LR},
 \ee
 where the LR subscript means that $u$ should be replaced by the value that maximizes the function 
 $u^2 (1-2u) \left( 1+ f^{\rm 2PM}(u) \right)$. In the present (2PM-accurate) case, this means $u^{\rm 2PM}_{\rm LR}=\frac13$ so that
 \be
 \left[\frac{d\e}{dj}\right]^{2\rm PM} \overset{\rm HE}{=} \sqrt{ \frac1{27} \left(1+ \frac{15}{54}\right)} \approx 0.217543,
 \ee
 corresponding to a non-rescaled Regge slope of
 \be
 \frac{ds}{dJ}= \frac2G \frac{d\e}{dj} \overset{\rm HE}{\approx} \frac{0.435087}{G}\,.
 \ee
 The main interest of this result is not its numerical value (which is likely to be significantly modified
 by higher PM effects; see below), but the {\it independence} of this HE slope on the mass ratio.
 Indeed, from the HE results discussed in Section \ref{sec:HE1}, the result \eqref{slope2PM} generalizes
 to higher PM orders with exactly the same final expression, as given on the rhs of Eq. \eqref{slope2PM},
 but with a correcting function $f(u)$ modified by higher powers of $u$. 
 For instance, if we use the current Amati-Ciafaloni-Veneziano-based knowledge of $f(u)$, namely the value
 $f^{\rm ACV}(u)$ given by the rhs of Eq. \eqref{fACV} (truncated to the $u^4$ level included), we find
 a maximum value of $ u^2 (1-2u) \left( 1+ f^{\rm ACV}(u) \right)$ equal to $0.129587$ (reached for
 $u_{\rm LR}^{\rm HE}=0.413696$. This corresponds to
 a $\nu$-independent HE slope equal to 
 \be
 \frac{ds}{dJ}= \frac2G \frac{d\e}{dj} \overset{\rm HE}{\approx} \frac{0.719964}{G}\,.
 \ee
 Let us recall that an extremely rotating (Kerr) black hole  has a total mass-energy satisfying
 \be \label{BHslope}
 E_{\rm extreme \, BH}^2 = \frac{J}{G}\,,
 \ee
 formally corresponding to a Regge slope equal to $1/G$.
 
 Our results above mean that if we form a binary system by bringing together (in the c.m. frame) two high energy
 particles so that they hold, under their mutual (conservative) gravitational attraction, on an (unstable) circular orbit,
 they will have a total angular momentum related to the squared energy by a relation of the type
 \be \label{reggeC}
  \E^2 \overset{\rm HE}{=} C \frac {J}{G}\,,
 \ee
 with a universal, $\nu$-independent numerical constant $C$ of order one. [Seen from this perspective, the
 $\nu$-independence of $C$ is natural because the rest-mass contributions of the two objects become
 irrelevant in the HE limit.]
 The PM perturbative estimates above suggest that $C$ is smaller than 1 (though the fact \eqref{BHslope}
 suggests that $C$ might end up being equal to 1).
 
\begin{figure}
\includegraphics[scale=0.7]{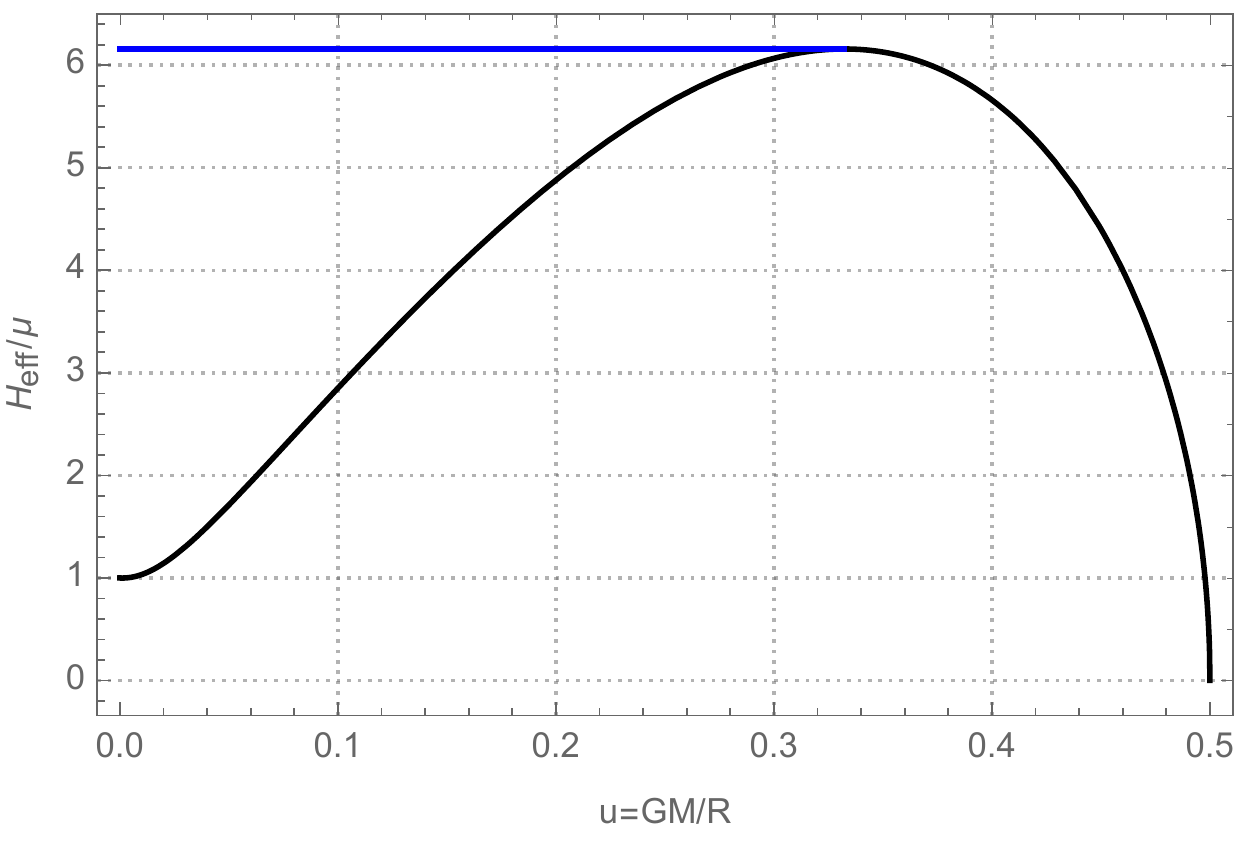}
\caption{\label{figradialpotential}
2PM-accurate, equal-mass ($\nu=\frac14 $) rescaled effective Hamiltonian $\hHf$ as a function of the inverse radial variable $u=GM/R$, for the rescaled angular momentum $j=30$. Note that radial infinity is at $u=0$ on the left. The horizontal line indicates the critical value of the effective energy for which the two-body system would end up (in absence of dissipation) in an infinite whirl motion.}
\end{figure}

 Using Eqs. \eqref{jbp}, \eqref{ecm}, we also deduce from Eq. \eqref{reggeC} that the critical impact parameter
 (in absence of dissipation) leading to  collapse, rather than scattering, in a HE collision 
 (see Fig. \ref{figradialpotential}) is equal to
 \be \label{bc}
 b_c \overset{\rm HE}{=} \frac{2\, G \E}{C}\,.
 \ee
 Several different lines of work have tried to estimate the value of $b_c$, see, e.g., 
 \cite{Eardley:2002re,Yoshino:2002tx,Giddings:2004xy,Amati:2007ak}. The construction of Ref. \cite{Eardley:2002re} 
 yielded the inequality $b_c \geq 3.219 \frac{G\E}{2}$, corresponding, via Eq. \eqref{bc}, 
 to $C \leq \frac{4}{3.219}\approx 1.243$.
 The analytical estimate of \cite{Amati:2007ak} corresponds to $C=2^{1/2}3^{-3/4}\approx 0.6204$.
 
 It would be interesting to perform simulations of  high-energy scattering of black holes to determine
 the value of the constant $C$. We tried to use the few existing simulations of high-energy scattering
 black hole encounters \cite{Shibata:2008rq,Sperhake:2012me} to determine the value of $C$.
 The idea is to focus on black hole motions of the asymptotically zoom-whirl type, 
 corresponding (in the Hamiltonian EOB representation)
 to an effective particle coming from infinity with a  high angular momentum $j \gg 1$ whose energy is just equal to 
 the maximum of the Hamiltonian (for the given value of $j$) so that the particle ends up, in the infinite future,
 on the (unstable) top of the Hamiltonian. This is illustrated (for the 2PM-Hamiltonian) in Fig. \ref{figradialpotential}.
 
 The problem, however, is that numerical simulations are studying
 dissipative motions. A cure for this problem was indicated in Ref. \cite{Bini:2012ji}, and was used (for slow
 black hole encounters) in Ref. \cite{Damour:2014afa}: it consists in subtracting the energy and angular momentum
 lost to gravitational radiation during the incoming motion, and to consider that the subtracted energy and
 angular momentum, $\e^{\rm in} - \e^{\rm rad}$,  $j^{\rm in} - j^{\rm rad}$, estimate the energy
 and angular momentum of the corresponding asymptotically-whirling motion of a {\it conservative} 
 binary motion. [There would also be the issue of taking care of the mass-energy absorbed by the
 black holes up to the moment of the first whirl.] Using some data, for  given
 in Ref. \cite{Shibata:2008rq} (and neglecting the effect of the absorbed mass-energy) for their highest
 velocity encounter ($v=0.9 \, c$), we found the rough estimate $C^{\rm num} \sim 0.9$. This estimate
 is consistent with our conclusions above. Clearly, new, higher-energy simulations, including estimates
 of gravitational radiation losses during the incoming motion, are needed to get any firm conclusion about
 the numerical value of $C$.
 
 Let us complete this Section by discussing several other consequences of the HE behavior of the PM-expanded
 Hamiltonian studied above.
 
 The first interesting consequence is the impossibility of transforming, in an  exact way, the 2PM
 Hamiltonian \eqref{Hf2PM} in a DJS-type gauge. We recall that Ref. \cite{Damour:2000we} has
 shown that it is possible, to all orders in the PN expansion, to find a PN-expanded canonical transformation
 such that the post-Schwarzschild term $Q$ in the EOB effective mass-shell, Eq. \eqref{massshellgen}, depends
 on quartic and higher powers of momenta only through the radial momentum $P_R$. In this DJS gauge the energetics
 of circular orbits is packed in the sole EOB radial momentum ${\bar A}( \bu; \nu) \equiv - g^{\rm eff}_{00}({\bar R})$.
 More precisely, the gauge-invariant energetics $\e(j)$ of circular orbits in DJS gauge 
 (with $\bu\equiv GM/(c^2 {\bar R})$) is obtained by eliminating $\bu$ between the two equations
 \be \label{hHDJScirc}
{\widehat H}^{2 \,\rm circ}_{\rm eff \, DJS}(\bu,j) = {\bar A}( \bu; \nu) (1+ j^2 \bu^2) \equiv {\bar A}( \bu; \nu) + j^2 {\bar B}( \bu; \nu)\,,
\ee
and
\be \label{DJScirc}
0=\frac{\partial {\widehat H}_{\rm eff \, DJS}^{2 \,\rm circ}(\bu,j ; \nu)}{\partial \bu}= {\bar A'}( \bu; \nu) + j^2 {\bar B'}( \bu; \nu) \,.
\ee
Here, we have introduced the notation ${\bar B}( \bu; \nu) \equiv \bu^2 {\bar A}( \bu; \nu)$ (which should not be
confused with the use of the letter $B$ to denote $g^{\rm eff}_{RR}$), and used a prime to denote the $\bu$
derivative.

Henceforth we consider the sequence of circular orbits, i.e. the solutions of the two equations \eqref{hHDJScirc},
\eqref{DJScirc}. In principle, all quantities can be considered as functions of $j$, or $j^2$, along the latter sequence 
(modulo the consideration of the two branches illustrated in Fig.~\ref{figej}).
For brevity, we do not add a superscript ``circ"  along the latter sequence. It is easily seen,
by differentiating \eqref{hHDJScirc}, that along circular orbits we have $ d\e^2(j^2)= {\bar B}( \bu(j^2); \nu) dj^2$.
Therefore, given the gauge-invariant functional link $\e(j)$, or $\e^2(j^2)$, we can recover the value of $ {\bar B}( \bu(j^2); \nu)$ along the circular sequence via
\be \label{BDJS}
{\bar B}(j^2) = \frac{d \e^2(j^2)}{dj^2} \,.
\ee
Inserting this result in Eq. \eqref{hHDJScirc} allows one to get also the value of ${\bar A}( \bu(j^2); \nu) $, namely
\be \label{ADJS}
{\bar A}(j^2) = \e^2(j^2) - j^2\frac{d \e^2(j^2)}{dj^2} \,.
\ee
Finally, in view of the definition  ${\bar B} \equiv \bu^2 {\bar A}$, we also get the value of $\bu^2(j^2)$, namely
\be\label{u2DJS}
\bu^2(j^2) =\frac{{\bar B}(j^2)}{{\bar A}(j^2)}=\frac{d \e^2(j^2)/dj^2}{\e^2(j^2) - j^2d \e^2(j^2)/dj^2}\,.
\ee
The above set of equations allows one to construct, in a parametrized way, the value of the 
DJS-gauge function ${\bar A}(\bu)$  from the sole knowledge of the gauge-invariant function $ \e^2(j^2)$.
At face value, it seems to give a non-perturbative (i.e. non PN-expanded) proof of the fact that one can
always encode the full circular energetics in the DJS-gauge function ${\bar A}(\bu)$. However, this reconstruction
is meaningful only if the quantity $\bu^2$ defined by Eq. \eqref{u2DJS} remains {\it finite and positive}
along the sequence of circular orbits. [One should additionally worry
about monotonicity issues.] 

We have applied the above reconstruction procedure to the energy curve defined by the  2PM-accurate Hamiltonian,
and represented above in Fig. \ref{figej}.
While a numerical calculation of ${\bar B}(j^2 ; \nu)$ from Eq. \eqref{BDJS} (i.e. essentially
a study of the slope of the curve in Fig. \ref{figej}) leads to an apparently acceptable, and positive, result,
the numerical calculation of ${\bar A}(j^2 ; \nu)$ from Eq. \eqref{ADJS} defines a quantity ${\bar A}$
which, for any non-zero value of $\nu$, changes sign near the LR (i.e. for large enough values of $j^2$
along the upper branch of the $ \e^2(j^2)$ curve). For instance, when $\nu=\frac14$, ${\bar A}$ 
vanishes around the 2PM-gauge $u$-parameter $u_*\approx 0.329806538$,
corresponding to $j_*\approx 14.8769$. Correspondingly, the quantity $\bu^2(j^2 ; \nu=\frac14)$ 
computed from Eq. \eqref{u2DJS}, which was positive along the stable branch and the beginning
of the unstable branch ($j^2<j_*^2$), becomes infinite at $j_*^2$, before becoming negative when getting
closer to the LR, i.e. when $j^2> j_*^2$. This result shows that there {\it does not exist} an exact canonical
transformation allowing one to transform the 2PM-accurate Hamiltonian into the DJS gauge. It also shows
(in confirmation of the findings of Ref. \cite{Akcay:2012ea}) that the obstruction to the construction of
a DJS gauge occurs, when seen  in phase space, only for large values of both the energy and the angular momentum.
More discussion about this below.
 
Finally, let us discuss the predictions made by the PM Hamiltonians of the type \eqref{Hf2PM} concerning the
behavior of Detweiler's redshift function near the LR. We recall that Detweiler \cite{Detweiler:2008ft} emphasized the
usefulness of considering, along the sequence of circular orbits of a two-body system, the gauge-invariant
function $z_1(x)$ (to which one can add $z_2(x)$), where $z_a=\left[ ds_a/dt \right]^{\rm reg}$ 
(with $a=1,2$) is the regularized value of the redshift along the worldline of the mass $m_a$ (with $m_1< m_2$,
and, in the SF case considered by Detweiler, $m_1 \ll m_2$). In our analytical PM estimates below, we use the results of 
\cite{LeTiec:2011ab,LeTiec:2011dp} to compute the redshift variables (along circular orbits) by means of a
partial derivative with respect to the rest-masses
\be \label{firstlaw}
z_a= \frac{\partial E_{\rm real}^{\rm circ}(J, m_a)}{\partial m_a} \,.
\ee

The parameter $x$ wich is generally used as gauge-invariant argument of
$z_a$ is the dimensionless frequency parameter
\be
x \equiv \left(\frac{G M \Omega}{c^3}  \right)^{2/3}\,,
\ee
where
\be
\Omega=\frac{  \partial E_{\rm real}^{\rm circ}(J, m_a)}{\partial J}\,,
\ee
 is the orbital frequency. [One often replaces, in SF studies, $x$ by
$y \equiv \left(G m_2 \Omega/c^3 \right)^{2/3}$, but we prefer here to use the 
$1 \leftrightarrow 2$-symmetric argument $x$.]

Our first result is that the parameter $x$ is actually a bad argument because it is {\it not monotonic} along
the sequence of circular orbits. This actually is already true at the (improved) 1PM level, and is a direct consequence
of one of the basic building blocks of EOB theory. Indeed, the (exact \cite{Damour:2016gwp}) EOB energy map
\eqref{f} shows that the orbital frequency is given by
\be \label{om}
\Omega=\frac{d E_{\rm real}}{dJ}= \frac{\Omega_{\rm eff}}{h},
\ee
where we recall that
\be
h \equiv \frac{\E}{M}= \sqrt{1+ 2\nu (\e-1)}\,,
\ee
and
\be \label{omf}
\Omega_{\rm eff} \equiv \frac{d {\mathcal E}_{\rm eff}}{dJ}\,.
\ee
By definition, $\e$, and therefore $h$ tends to infinity as one approaches the LR\footnote{Here, we consider
the exact (conservative) two-body LR, corresponding to an (unstable) ultrahigh-energy binary orbit.}
If we start by considering the 1PM approximation, i.e. the effective Hamiltonian of a particle
of mass $\mu$ in a Schwarzschild metric of mass $M$, we have the well-known result (see, e.g., 
\cite{Buonanno:1998gg,Damour:2009sm}) that the parameter $u=GM/R$ is a monotonic parameter along the
sequence of circular orbits (with $0<u<\frac13$), in terms of which one has
\be
\frac{GM \Omega_{\rm eff}^{\rm 1PM}}{c^3} = u^{3/2}\,,
\ee
and
\be
\e^{\rm 1PM}=\frac{1-2u}{\sqrt{1-3u}} \,;\, j^{\rm 1PM}=\frac{1}{\sqrt{u(1-3u)}}\,.
\ee
If we (generally) define an ``effective" frequency parameter as
\be
x_{\rm eff} \equiv \left(\frac{G M \Omega_{\rm eff}}{c^3}  \right)^{2/3} \equiv h^{2/3} \, x\,,
\ee
we have simply $x_{\rm eff}^{\rm 1PM}=u$ and therefore
\be
x^{\rm 1PM}=\frac{u}{\left[1+ 2\nu (\e^{\rm 1PM}-1) \right]^{1/3}}\,.
\ee
The latter result shows that $x^{\rm 1PM}$ tends to zero as one approaches the LR. Therefore the curve
$x^{\rm 1PM}(u)$, which starts at the origin as $x^{\rm 1PM}(u) = u+O(\nu u^2)$ for small $u$,
then turns back towards zero as $u$ nears $u=\frac13$ while ranging over the interval $0<u<\frac13$.
This shows that the link $x \to z_a$ does not define a function. It also shows that, even at the 1PM approximation,
the SF expansion of the formal link $z_a(x)$ will be necessarily singular at the LR.

\begin{figure}
\includegraphics[scale=0.7]{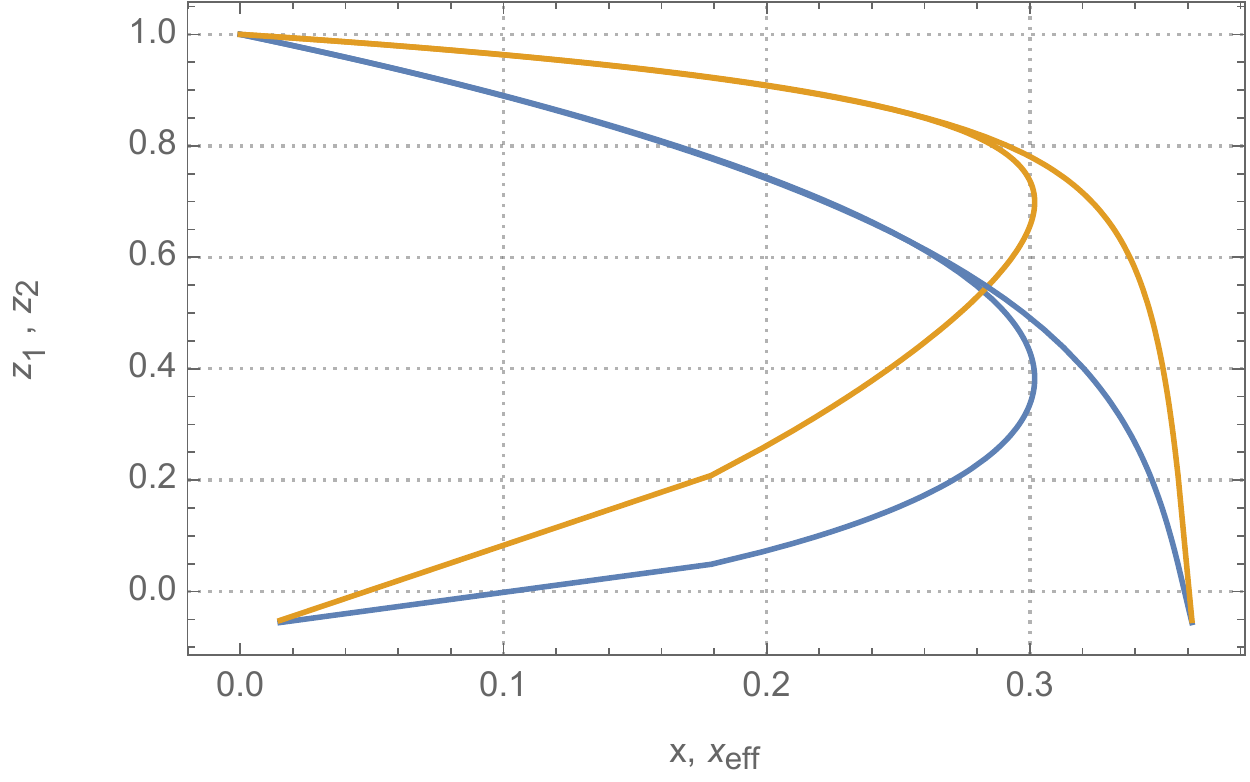}
\caption{\label{figz1z2xxf}
Graphs of the relations (for $\nu=0.2$ and the 2PM Hamiltonian) between the two redshifts $z_a$ and either the usual frequency parameter $x$ (leading to
the two curves that turn back towards the left) or the EOB-motivated effective frequency parameter $x_{\rm eff}$.
Only the latter choice defines functions $z_a(x_{\rm eff};\nu)$.}
\end{figure}

We therefore propose to replace Detweiler's original reshift function $z_a(x)$ by the EOB-motivated 
functional link $x_{\rm eff} \to z_a$, which defines two good functions at the 1PM level. From numerical
simulations, it seems that the functions $z_a(x_{\rm eff};\nu)$ still define good functions at the 2PM level.
This is illustrated in Fig. \ref{figz1z2xxf} which compares (for the 2PM Hamiltonian, and for $\nu=0.2$) the two functions $z_a(x_{\rm eff};\nu)$, to the parametric curves representing the links $z_a(x;\nu)$.

The two functions $z_a(x_{\rm eff};\nu)$ are ordered as expected from
the large-mass-ratio limit, i.e. $ z_1(x_{\rm eff};\nu) \leq z_2(x_{\rm eff};\nu) \leq 1$. [We will comment later on
the limiting values of $z_1, z_2$ at the LR.] 
Let us emphasize that the value of $x_{\rm eff}$ at the LR (i.e. at infinite energy) is finite,
and that the function $x_{\rm eff}^{\rm LR}(\nu)$ monotonically increases with $\nu$ from $\frac13$ when $\nu=0$
to $x_{\rm eff}^{\rm LR}(\frac14) \approx 0.3617$. This corresponds to a fractional increase (when passing from
$\nu=0$ to $\nu=\frac14$) in the
effective orbital frequency at the LR of $\sim 13$\%. Remember again that, by contrast,
 the real orbital frequency \eqref{om} at 
the LR vanishes for all non-zero values of $\nu$. 

It would be interesting to try to extend the existing direct numerical estimates of the functions
$z_a(x)$, recently obtained in Ref. \cite{Zimmerman:2016ajr} (which were limited to the range $ GM \Omega \lesssim 0.1$,
corresponding to $ x\approx x_{\rm eff} \lesssim 0.215$), to the full range considered here, i.e. 
up to the LR. This is, however, a challenging task for several reasons. On the one hand, we are discussing here
the conservative dynamics while numerical simulations give access to the dissipative dynamics. [It
was, however, indicated above how to correct for that when discussing the energetics.] On the other hand, the
formal dynamical LR discussed here for two point particles might be preceded, when realizing these particles
as black holes, by the coalescence of the two horizons. [Indeed, the fact that the orbital frequency along
the sequence of conservative circular motions reaches a maximum before the LR (where it formally vanishes)
is reminiscent of the EOB prescription (along low-energy, post-LSO plunging motions) to define merger as
the moment where the orbital frequency reaches a maximum.]

Let us end this Section by emphasizing the link between the HE Regge behavior \eqref{reggeC} and the LR
behavior of the redshifts $z_a$. First, we note that the leading-order HE relation \eqref{reggeC} predicts that
$E_{\rm real}$ is only a function of $J$, without any dependence on the two masses $m_a$. This would seem
to imply, according to the first law \eqref{firstlaw}, that the redshifts $z_a$ must tend to zero at the LR.
However, one must take into account the next-to-leading-order (NLO) contribution to the Regge-type relation 
\eqref{reggeC}. A look at the 2PM Hamiltonian \eqref{Hf2PM} (considered in the circular case, $p_r=0$), 
shows that the ratio $\hHf^2/j^2$ tends, in the HE limit, to a function of $u$ modulo $j$-dependent
fractional corrections, namely
\be \label{NLOHf2byj2}
\frac{\hHf^2}{j^2}= B_{\rm HE}(u) \left[ 1 + O\left( \frac1{ (2\nu j)^{1/2} (u^2(1-2u))^{1/4} }  \right)  \right] \,,
\ee
where
\be
B_{\rm HE}(u) \equiv u^2 (1-2u) (1+f(u)) \,.
\ee
Here, we have indicated only the leading-order fractional corrections in inverse powers of $j$.

Extremizing the rhs of \eqref{NLOHf2byj2} with respect to $u$ we recover, at leading order the result \eqref{slope2PM}.
But we also get the additional information that the fractional correction to the slope \eqref{slope2PM} is
(modulo numerical factors) of order $\sim 1/(\nu j)^{1/2}$. Converting this information in terms of the Regge-type
relation \eqref{reggeC}, we see that the NLO version of the HE Regge relation is
\be \label{NLOreggeC}
  \E^2 = C \frac {J}{G} \left[1 + \beta (m_1+m_2) \sqrt{\frac{ G}{J} } + O\left(\frac1{J}\right)\right] ,
 \ee
 or, equivalently,
 \be
 \E = \sqrt{C} \sqrt{\frac {J}{G}}  + \frac12 \beta \sqrt{C} (m_1+m_2)  + O\left(\frac1{\sqrt{J}}\right) .
 \ee
 Using \eqref{firstlaw}, we then deduce that the (formal) LR limits of the redshifts are finite and equal to
 \be
 z_1^{\rm LR} =  z_2^{\rm LR} =\frac12 \beta \sqrt{C}.
 \ee
 When using, as is, the 2PM-accurate Hamiltonian, one finds that the value of $\beta$ is (small and) negative
 (as exhibited in Fig. \ref{figz1z2xxf}). Evidently, as was already the case for the
 numerical value of the  leading-order Regge slope $C$, we expect that the numerical value of $\beta$
 will be significantly modified by higher-order PM contributions. One would have naively expected a 
 vanishing value of $z_a$ at the LR. When evaluating $z_a$ for black holes (rather than point masses,
 which involve a regularization of $z_a$), the redhifts are evaluated as a ratio of two (a priori positive)
 surface gravitities \cite{Zimmerman:2016ajr}. We would then expect $z_a$ never to become negative.
 Only future work (and a determination of the higher PM versions of the Hamiltonian) will be able to
 decide whether the correct value of $\beta$ is positive (or zero). If the analytical estimates of the type
 presented here continue to produce a negative value of $\beta$ this might signal that we are trusting
 our analytical description (EOB, as well as the first law) beyond its physical domain of applicability.
 [For instance, one might have to stop using the description at the threshold where the smallest
 redshift vanishes.]

\section{Towards translating  quantum gravitational scattering amplitudes into classical dynamical information.}

We have already shown above how to translate the HE scattering results of Amati, Ciafaloni and Veneziano 
into information about the structure of the EOB effective Hamiltonian. But our task had been facilitated 
by the fact that Amati, Ciafaloni and Veneziano had already translated their quantum results in 
terms of a quasi-classical eikonal approximation.
In this section, we wish to discuss how to relate the classical dynamical information contained in the EOB Hamiltonian
to  the perturbative quantum 2-to-2 gravitational scattering amplitude, given by
a Born-type, coupling-constant expansion of the form
\be \label{bornexp}
\cM(s,t) = \cM^{\left( \frac{G}{\hbar}\right)}(s,t)  + \cM^{\left(\frac{G^2}{\hbar^2}\right)}(s,t)  + \cdots\,,
\ee
where $s = -(p_1+p_2)^2$ and $t=-(p'_1-p_1)^2$ are Mandelstam variables\footnote{We consider the scattering
of scalar particles of mass $m_1$ and $m_2$, from the ingoing state $| p_1p_2\rangle$ to the outgoing state 
$| p'_1p'_2\rangle$.} and where each term is proportional to a power of  $G/\hbar \equiv M_P^{-2}$. [Here, $M_P$ denotes the Planck mass;
we recall that we use $c=1$, while we keep $G$ and $\hbar$.]

The first Born approximation is (see, e.g., Ref. \cite{Kabat:1992tb})
\be \label{born1}
 \cM^{\left( \frac{G}{\hbar}\right)}(s,t) = 16 \pi \frac{G}{\hbar} \,\frac{2 \, (p_1\cdot p_2)^2 - p_1^2 \, p_2^2}{-t}\,.
\ee
Note that $\cM^{\left( \frac{G}{\hbar}\right)}(s,t)$ is positive for the real scattering kinematics (with, notably, $s>0$ and $t<0$.] We use the sign convention where the scattering matrix is 
\be
\langle p'_1p'_2|  S | p_1p_2\rangle={\rm Identity}+ i (2\pi)^4 \delta^4(p_1+p_2-p'_1-p'_2) \frac{\cM}{N},
\ee 
with a dimensionless Lorentz-invariant amplitude $\cM$ and a (state-normalization-related) positive numerical factor 
$N$, given (when using the state normalization $\langle p'|p\rangle = (2\pi)^3 \delta^3(\p-\p')$) by $N= (2E_1)^{1/2} (2E_2)^{1/2} (2E'_1)^{1/2} (2E'_2)^{1/2}$.
With this sign convention $\cM$ is proportional, in the case of potential scattering, to the usual nonrelativistic
outgoing scattering amplitude $f(\Omega)$ measuring the coefficient of the scattered, outgoing wave,
$\psi^{\rm scatt} = f(\Omega) e^{i kr}/r$, where $\Omega$ is the scattered direction on the sphere.
The first Born approximation of $f$ is proportional to the matrix element of {\it minus} the potential,
so that $f$ is positive for an attractive interaction potential.

If we consider, for orientation, a case where $s \sim -t$ and where the momenta are either comparable to or large
with respect to the rest masses, we have the order of magnitude $ \cM^{\left( \frac{G}{\hbar}\right)}(s,t) \sim E^2/M_P^2$. The second term in the Born-type expansion \eqref{bornexp} will then be
$ \cM^{\left( \frac{G^2}{\hbar^2}\right)}(s,t) \sim E^4/M_P^4$. We are then talking about an expansion valid
for $E \ll M_P$. Clearly, we cannot directly apply this expansion to the physical case of two black holes or two
neutron stars. We shall see how to bypass this problem by quantizing the EOB Hamiltonian. 

As already mentioned in \cite{Damour:2016gwp}, and as is clear when comparing Eq. \eqref{born1}
to Eq. \eqref{chiG}, there is a simple link between the $O(G)$ contribution to $\cM$ and the 1PM scattering angle.
 Our aim here is to relate the $O(G^2)$ scattering amplitude $ \cM^{\left( \frac{G^2}{\hbar^2}\right)}(s,t)$ to the 2PM scattering angle, and its corresponding 2PM Hamiltonian contribution \eqref{Hf2PM}. An a priori stumbling block
 in this task is the well-known fact that,  {\it a priori}, the domain of validity of the Born expansion is 
 $G E_1 E_2/( \hbar v) \ll 1$ (where $v$ is a characteristic relative velocity), while the domain of validity of
 classical scattering is the reverse condition, namely $G E_1 E_2/( \hbar v) \gg 1$ \cite{LandauQM} . 
 The link between the two different $O(G)$
 results then appears  to be accidental, and due to the fact that the Born approximation for Coulomb scattering happens
 to yield the exact differential cross section. To bypass this problem we propose to consider the quantum scattering
 defined by quantizing the EOB Hamiltonian dynamics. 
 
 For simplicity, we restrict ourselves here to the 2PM dynamics (keeping only $q_2$ in Eq. \eqref{Hf2PM}).
 If we use the rescaled variables \eqref{rescaled} the 2PM mass-shell condition reads
 \be
 g_0^{\mu \nu} p_\mu p_\nu + 1 + \hQ=0\,,
 \ee
 where $g_0$ is the Schwarzschild metric. Let us use isotropic coordinates for $g_0$, i.e.
 \be
 ds^2_0= - {\bar A}(\bu) dt^2 + {\bar B}(\bu) \left( d \br^2 + \br^2  (d \theta^2 + \sin^2 \theta d \varphi^2) \right),
  \ee
 with $\bu \equiv 1/\br$ and
 \be
 {\bar A}(\bu)= \left(\frac{1- \frac12 \bu }{1+ \frac12 \bu}\right)^2 \, ;\,  {\bar B}(\bu)= \left(1+ \frac12 \bu   \right)^4 .
 \ee
 Using Cartesian coordinates $x^i ={\bf x} $ linked in the usual way to $\br, \theta, \varphi$, and denoting the
 covariant  momenta $p_i$ as $ \p$, the 2PM-accurate (rescaled) mass-shell condition reads 
 (using $u=\bu+ O(\bu^2)$, so that $u^2=\bu^2+ O(\bu^3)$)
 \be
 0=- \frac{ \e^2 }{{\bar A}} + 1 + \frac{\p^2}{ {\bar B} } + \bu^2 q_2(\e) + O(\bu^3) \,,
 \ee
 or, equivalently, multiplying by ${\bar B}$ and using $ \bu^2 {\bar B} =\bu^2 + O(\bu^3)$,
 \be
  0=- \frac{  {\bar B} }{{\bar A}}  \e^2+  {\bar B}  + \p^2+ \bu^2 q_2(\e) + O(\bu^3) \,.
 \ee
 This yields a 2PM-accurate mass-shell condition of the form
 \be \label{2PMmassshell}
 \p^2 = p_{\infty}^2 + {\overline W}(\bu) =  p_{\infty}^2 + w_1 \bu + w_2 \bu^2 +   O(\bu^3) \,,
 \ee
 with the following energy-dependent coefficients
 \bea \label{defpinfw1w2}
 p_{\infty}^2 &=& \e^2-1 ,\nonumber \\
 w_1 &=& 2 (2 \e^2-1) ,\nonumber \\
 w_2 &=& \frac32 \frac{5 \e^2-1}{h(\e)}.
\eea
In the last coefficient $h$ denotes the rescaled real energy, Eq. \eqref{hEj}. Note that the $O(\bu^2)$ term
in the mass-shell condition \eqref{2PMmassshell} has resulted from the sum of the original 2PM term
$ \bu^2 q_2(\e) = \bu^2 \frac32 (5 \e^2-1)(1-1/h)$ and of a term coming from the expansion
in powers of $\bu$ of the potential-like terms $- \frac{  {\bar B} }{{\bar A}}  \e^2+  {\bar B}$ that
have exactly cancelled the term proportional to 1, instead of $1/h$, in $ \bu^2 q_2(\e) $.

We can now straightforwardly quantize the PM-expanded  mass-shell condition \eqref{2PMmassshell}.
Remembering the rescalings of ${\bf x}$ and $\p$, their commutation relation reads
\be
[x^i , p_j] = i \hh \,\delta^i_j \; {\rm where} \; \hh \equiv \frac{\hbar}{GM\mu} = \frac{\hbar}{G m_1 m_2}.
\ee
Note that $\hh$ is dimensionless and is essentially equal, when considering a mildly relativistic ($v \sim 1$) scattering
with $E_1 \sim E_2 \sim m_1 \sim m_2$, to the {\it inverse} of the expansion parameter of the Born approximation.

Considering a fixed energy, we get the following time-independent Schr\"odinger equation
\be \label{schro}
- \hh^2 \Delta_{\bf x} \psi({\bf x}) = \left[  p_{\infty}^2 + \frac{w_1}{\br} + \frac{w_2}{\br^2} +   O\left(\frac1{\br^3}\right)    \right]\psi({\bf x}) \,.
\ee
One should remember that $\frac1{\br}= \frac{GM}{{\bar R}}$ is of order $O(G)$.
We then see that, finally, in the EOB formulation (in isotropic coordinates), the quantum scattering of two (scalar) particles is
described by the scattering of a (scalar) effective particle on an energy-dependent potential which is
the sum, at the 2PM approximation, of a Newton $GM/{\bar R}$ potential and of a correcting $\left(GM/{\bar R}\right)^2$ term.

It is interesting to note that though we are discussing relativistic scattering, the EOB formulation has allowed us
to reduce the computation of the scattering amplitude to a nonrelativisticlike potential scattering problem.
Using standard results from quantum potential scattering \cite{Messiah}, and denoting the asymptotic 
plane waves as\footnote{Beware that $\vk$ and $\x$ are rescaled versions of the usual wave  and position vectors.}
\be
\varphi_a = e^{i  \vk_a \cdot \x } \; ; \; \varphi_b = e^{i  \vk_b \cdot \x } \,,
\ee
where the label $a$ refers to the ingoing state and the label $b$ to the outgoing one, the stationary  retarded-type 
solution of the scattering equation \eqref{schro}, say $\psi^+_a$, describing a state $|\vk_a\rangle$ in the infinite past,
has the following structure at large distances (with $r=| \x|$)
\be
\psi^+_a  \underset{r \to \infty}{\approx} e^{i  \vk_a \cdot \x } + f^+_{\vk_a}(\Omega) \frac{e^{i k r}}{r}\,.
\ee
In this formulation, the quantity $f^+_{\vk_a}(\Omega_b)$ (which differs from $\cM(s,t)$ only by 
some normalization factor) measures the c.m. scattering amplitude in the outgoing
direction $\Omega_b = \vk_b/k$. The conserved norm of the wave vector,  $k= |\vk_a| = |\vk_b|$, is related
to $ p_{\infty}$, Eq. \eqref{defpinfw1w2}, via
 \be \label{pinfvsk}
p_{\infty} = \hh \, k \,.
\ee
The scattering amplitude is given by
\be
f^+_{\vk_a}(\Omega_b) = +\frac1{4\pi \hh^2} \langle\varphi_b| \overline W |\psi^+_a \rangle\,,
\ee
where
\be
 \overline W = + \frac{w_1}{\br} + \frac{w_2}{\br^2} +   O\left(\frac1{\br^3}\right) \,,
\ee
is {\it minus} the potential in the Schr\"odinger equation \eqref{schro}. In other words, we consider as
Hamiltonian $H = \frac{\p^2}{2m}+ V= - \hh^2 \Delta_{\bf x} - {\overline W}$ with $m=\frac12$,
 $\p= \frac{\hh}{i} \frac{\partial}{\partial \x}$, and, asymptotically $\p_a = \hh \,\vk_a$ and $\p_b = \hh \,\vk_b$.
[Our conventions maximize the number of plus signs in the relevant equations.]

The first-order Born (B1) approximation is
\bea \label{B1}
f^{+ \, \rm B1}_{\vk_a}(\Omega_b) &=&  +\frac1{4\pi \hh^2} \langle\varphi_b| \overline W |\varphi_a \rangle \nonumber \\
 &=&  +\frac1{4\pi \hh^2} \int d^3\x e^{- i {\bf q} \cdot \x} \overline W  \,,
\eea
where 
\be \label{qvsk}
{\bf q} = \vk_b - \vk_a \; ; \; q = |{\bf q}| = 2 k \sin \frac{\theta}{2} .
\ee
Here, $\theta$ denotes the angle between $\vk_a$ and $\vk_b$, so that
the scattering amplitude $f^+$ is a function of $\theta$. The link between the dimensionless
quantity  $q$ and the physical c.m. momentum transfer $Q^{\rm c.m.} = \sqrt{-t}$ will be discussed below.

When reinstating the gravitational constant, the potential ${\overline W}$ is a sum 
 ${\overline W} = \sum_n w_n/\br^n$, with $ w_n/\br^n =O(G^n)$. The first-order Born
 approximation, Eq. \eqref{B1}, is then obtained by computing the Fourier transform of $1/\br^n$ potentials.
 These are obtained from the general formula (in space dimension $d$)
\be
{\cal F}^{(d)}\!\left[ \frac1{r^n}\right] \equiv \int d^d \x e^{- i \vk \cdot \x} \frac1{r^n} = \frac{C_n^{(d)}}{k^{d-n}}\; ,
\ee
 where
 \be
 C_n^{(d)} =\pi^{\frac{d}{2}} \frac{2^{\bar n} \Gamma(\frac12 {\bar n})}{\Gamma(\frac12 n)}\; ; \; {\rm with} \; {\bar n} \equiv d -n \,.
 \ee
 This general formula yields 
 \be
 {\cal F}^{(3)}\!\left[ \frac1{r}\right] =\frac{4 \pi}{k^2} \; ;  \; {\cal F}^{(3)}\!\left[ \frac1{r^2}\right]=\frac{2 \pi^2}{k} \,,
 \ee
 so that
 \be
 f^{+ \, \rm B1}_{\vk_a}(\vk_b)=\frac1{\hh^2} \left[ \frac{w_1}{q^2} + \frac{\pi}{2} \frac{w_2}{q}  \right]\,.
 \ee

 The second ($w_2$) term in this result is already of order $O(G^2)$, while the first one is $O(G)$.
 To obtain the scattering amplitude to the $O(G^2)$ accuracy, one a priori needs to consider the second-order
 Born approximation. However, only the Newtonianlike potential $w_1/\br$ contribution needs to be iterated
 to second order. The latter, second Born iteration is straightforwardly derived from considering the known,
 exact Coulomb scattering amplitude \cite{LandauQM}. This can be embodied in a correcting factor $F_C=e^{\delta_C}$ multiplying the
 $w_1$ contribution above. Finally, the $O(G^2)$-accurate scattering amplitude derived by quantizing the
 EOB effective Hamiltonian reads
 \be
  f^{+ \, \rm B1}_{\vk_a}(\vk_b)=\frac1{\hh^2} \left[ e^{\delta_C} \frac{w_1}{q^2} + \frac{\pi}{2} \frac{w_2}{q}  \right],
 \ee
 where
 \be
 \delta_C= i\frac{w_1}{2k\hh^2} \ln(\sin^2 \frac{\theta}{2})+ 2i \arg \Gamma\left(1-i \frac{w_1}{2k\hh^2}\right).
 \ee
 The exponent $\delta_C$ in the correcting factor $F_C=e^{\delta_C}$ is mainly imaginary, but has also
 a real part coming from its second term.
 
 The simplest way to use this result without worrying about the issue of the relative
 normalization between $\cM$ and $f^+$ is to consider the ratio of the contribution $\propto 1/q$ to
 the one $\propto 1/q^2$, namely
 \bea
 \frac{f^+_{(1/q)}}{f^+_{(1/q^2)}}&=& \frac{\pi}{2} \frac{w_2}{w_1}  e^{-\delta_C} \, q\nonumber \\
  &=& \frac{3\pi}{8}\frac{5\e^2-1}{2\e^2-1} \frac{ q }{h(\e)} +O(G^2) .
 \eea
 In order to express this result in terms of standard physical quantities, we need to convert
 the rescaled EOB momentum transfer $ q=|{\bf q}| = |\vk_b - \vk_a| $ in terms of the physical momentum transfer $Q^{\rm c.m.} = \sqrt{-t}$. This is achieved by first using the relation \cite{Damour:2016gwp}
 \bea
 \E P_{\rm c.m.} &=& \sqrt{(p_1.p_2)^2 - p_1^2 p_2^2} \nonumber \\
 &=& m_1 m_2 \, \sqrt{\e^2-1} = \mu M p_\infty
 \eea
 Using the definition \eqref{defpinfw1w2} of $p_{\infty}$, this yields
 \be
P_\infty^{\rm EOB} \equiv \mu \, p_\infty = \frac{\E}{M} P_{\rm c.m.}= h(\e) P_{\rm c.m.}\,.
 \ee
 As a second step, we use the link \eqref{pinfvsk} between $k$ and $p_{\infty}$,
together with Eq. \eqref{qvsk}, and the fact that the physical c.m. scattering angle $\chi$ is equal to the EOB one $\theta$ \cite{Damour:2016gwp}. This yields
\bea
q  &=& 2 \sin \frac{\theta}{2} \frac{p_{\infty}}{\hh} = 2 \sin \frac{\chi}{2}  \frac{h(\e) P_{\rm c.m.}}{\mu \,\hh}\nonumber \\
&=&   \frac{G M}{\hbar} h(\e) Q^{\rm c.m.} \,,
\eea
where we used $ \sqrt{-t} = Q^{\rm c.m.}=  2 \sin \frac{\chi}{2} P_{\rm c.m.}$, and $\hh=\hbar/(GM\mu)$.
We finally get
\be
\frac{f^+_{(1/q)}}{f^+_{(1/q^2)}} = \frac{3\pi}{8}\frac{5\e^2-1}{2\e^2-1} \frac{ G (m_1+m_2) \sqrt{-t} }{\hbar} +O(G^2).
\ee

 This ratio should be equal to the ratio $\cM_{(1/\sqrt{-t})}/\cM_{(1/(-t))}$, with
 \be
  \cM^{\left( \frac{G}{\hbar (-t)}\right)}(s,t) = 16 \pi \frac{G m_1^2 m_2^2}{\hbar} \,\frac{2 \, \e^2 - 1}{-t}\,.
 \ee
 There are several recent works who used modern amplitudes techniques to compute the full $O(G^2)$, one-loop,
 two-graviton exchange, contribution to gravitational scattering amplitudes $\cM$. See notably 
 \cite{Bjerrum-Bohr:2013bxa,Bjerrum-Bohr:2014zsa,Bjerrum-Bohr:2016hpa}.
 $\cM^{(G^2)}(s,t)$ contains
 several types of terms linked to various topologies of the reduced scalar diagrams associated with $\cM$. It is,
 however, possible (as discussed in Refs. \cite{Bjerrum-Bohr:2013bxa,Bjerrum-Bohr:2014zsa,Bjerrum-Bohr:2016hpa})
 to extract from $\cM^{(G^2)}(s,t)$ the pieces corresponding to the $1/(-t)$ and $1/\sqrt{-t}$
 terms discussed above, which we have seen to be directly connected with interaction terms in the classical effective Hamiltonian. [Beware of some misprints in Refs. \cite{Bjerrum-Bohr:2014zsa,Bjerrum-Bohr:2016hpa}: the relative sign of  $\cM^{(G^2)}(s,t)$
 and $\cM^{(G^1)}(s,t)$ should be changed.] However, the latter references consider limits where the two-body effects
we are interested in (with explicit dependence on $m_1$ and $m_2$)   disappear. The only exception I am aware of
 is an unpublished work in preparation \cite{Bjerrum-BohrVanHove} which seems to be in full agreement with our results here.
 I hope that the present investigation will prompt further work along these lines, and, notably a computation of
 the  $O(G^3)$, {\it two-loop} quantum scattering amplitude. Generalizing the calculations of this Section,
 one should be able to extract from $\cM^{(G^3)}(s,t)$ the 3PM contribution $u^3 q_3(\e)$ to the effective
 two-body Hamiltonian which would significantly improve our knowledge of classical high-energy gravitational
 interactions.
 
 \section{Conclusions}
 
 Having in mind the needs of the upcoming era of high signal-to-noise-ratio gravitational-wave
observations, we have derived the $G^2$-accurate, second post-Minkowskian (2PM)  effective one-body (EOB)
Hamiltonian description of the conservative dynamics of two gravitationally interacting bodies having an
arbitrary (possibly relativistic) relative velocity. This result, which generalizes our previous 1PM work, was
obtained from the 2PM c.m. scattering angle derived long ago by Westpfahl and collaborators.
We stressed the similarity between the classical PM perturbative expansion of the scattering angle to the
Feynman perturbative expansion of quantum scattering amplitudes (see Section \ref{sec2}). 
It would be interesting to study in more detail this similarity, and to see whether it could allow one to
translate some of the improved, modern  quantum amplitude techniques into 
corresponding, improved classical scattering computations.

The effective 2PM EOB Hamiltonian, Eq. \eqref{Hf2PM}, was found to have an interesting high-energy (HE) structure,
with many attendant physical consequences: (i) while confirming a previous finding about a singular HE behavior of the
self-force expansion of the two-body dynamics, it shows that the exact (non self-force-expanded) two-body 
Hamiltonian is regular in the HE limit; (ii) the HE regularity of the two-body EOB Hamiltonian can only be obtained
in certain phase-space gauges, which necessarily differ from the gauge standardly used in the current (low-energy)
versions of the EOB dynamics; (iii) in the HE limit, the values of the two rest masses become unimportant and this
allowed us both to connect our results with, and exploit, the HE scattering results of Amati, Ciafaloni and Veneziano,
and to make predictions about some 3PM and 4PM effects, and about the energetics of HE circular (and zoom-whirl) orbits. We notably found that
high angular momenta, high energy circular orbits exhibit, to leading order, a (rest-mass independent) 
linear Regge trajectory behavior, Eq. \eqref{reggeC}. 
Ways of testing these predictions by dedicated numerical simulations were indicated. See also Eq. \eqref{NLOreggeC}
for the next-to-leading-order correction to the leading HE linear Regge behavior \eqref{reggeC}.

Finally, we indicated a way to connect our classical results to the quantum gravitational scattering amplitude of two
particles. We urge amplitude experts to use the available, efficient techniques to compute the 2-loop scattering
amplitude of scalar masses. Higher-loop generalizations  of the massless two-loop amplitude result of
Amati, Ciafaloni and Veneziano would also be quite interesting. We leave to future work the use of
the quantum dynamics defined by the EOB Hamiltonian as a new handle on a quantum description
of gravitational collapse during two-body collisions.

In view of the effectiveness of the current formulations of the EOB dynamics, which have played an
important role in the data analysis of the recent LIGO-Virgo observations, there is no urgent need to
reformulate the EOB Hamiltonian along the lines suggested here. However, we think that the upcoming
era of high signal-to-noise-ratio might benefit from studying whether a Numerical-Relativity completion
of the type of new, PM-suggested phase-space gauge employed here leads to a more accurate description
of the last orbits of coalescing black holes.


\end{document}